\documentclass[fleqn,usenatbib]{mnras}

\usepackage{newtxtext,newtxmath}

\usepackage[T1]{fontenc}

\DeclareRobustCommand{\VAN}[3]{#2}
\let\VANthebibliography\thebibliography
\def\thebibliography{\DeclareRobustCommand{\VAN}[3]{##3}\VANthebibliography}

\usepackage{graphicx}	
\usepackage{amsmath}	
\usepackage{ulem}
\usepackage{subcaption}

\newcommand{\dtm}{dust-to-metals} 
\newcommand{\msun}{\,M$_{\odot}$} 
\newcommand{\mthc}{\,$M_{200\text{c}}$}
\newcommand{\rthc}{\,$R_{200\text{c}}$}

\newcommand{\hMpc}{{\ifmmode{,h^{-1}{\rm Mpc}}\else{$h^{-1}$Mpc}\fi}}
\newcommand{\hkpc}{{\ifmmode{,h^{-1}{\rm kpc}}\else{$h^{-1}$kpc}\fi}}
\newcommand{\hMsun}{{\ifmmode{\,h^{-1}{\rm {M_{\odot}}}}\else{$h^{-1}{\rm{M_{\odot}}}$}\fi}}

\newcommand{\Mstar}{{\ifmmode{,M_{*}}\else{$M_{*}$}\fi}}
\newcommand{\Mhalo}{{\ifmmode{\,M_{\rm halo}}\else{$M_{\rm halo}$}\fi}}
\newcommand{\ltsima}{$\; \buildrel < \over \sim \;$}
\newcommand{\gtsima}{$\; \buildrel > \over \sim \;$}
\newcommand{\lsim}{\lower.5ex\hbox{\ltsima}}
\newcommand{\gsim}{\lower.5ex\hbox{\gtsima}}

\newcommand{\gadget}{\textsc{gadget-4}}

\newcommand{\powderday}{\textsc{powderday}}

\labelformat{section}{~#1}
\labelformat{subsection}{Section~#1}
\labelformat{subsubsection}{Sec.~#1}
\labelformat{figure}{Figure~#1}
\labelformat{subfigure}{Figure~\thefigure #1}
\labelformat{table}{Table~#1}
\labelformat{equation}{Equation~#1}

\title[Dust in cold and warm dark matter]{Dust in cold and warm dark matter in a suite of zoom-in simulations of Milky Way analogues}

\author[A. J. Ussing]{
Adam J. Ussing,$^{1,2}$\thanks{E-mail: aussing@swin.edu.au}
Robert A. Mostoghiu Paun,$^{1,2}$
Darren Croton,$^{1,2}$\thanks{E-mail: dcroton@swin.edu.au}
Alan Duffy,$^{1,2}$
\newauthor{}
Chris Power,$^{3}$
and Celine Boehm,$^{2,4,5}$
\\
$^{1}$Centre for Astrophysics and Supercomputing, Swinburne University of Technology, PO Box 218, Hawthorn VIC 3122, Australia\\
$^{2}$ARC Centre of Excellence for Dark Matter Particle Physics (CDM), Melbourne 3000, Australia\\
$^{3}$International Centre for Radio Astronomy Research (ICRAR), M468, University of Western Australia, 35 Stirling Hwy, Crawley, WA 6009, Australia\\
$^{4}$School of Physics, The University of Sydney, NSW 2006, Australia\\
$^{5}$School of Physics and Astronomy, The University of Edinburgh, UK\\
}
\date{Accepted 2026 October 21. Received 2026 October 1; in original form 2026 February 19}

\pubyear{\the\year{}}

\begin{document}
\label{firstpage}
\pagerange{\pageref{firstpage}--\pageref{lastpage}}
\maketitle

\begin{abstract}
We use a suite of hydrodynamical zoom-in simulations of Milky Way analogues to investigate how previously identified trends in dust properties of galaxies can be utilised to infer the underlying dark matter model. We use a sample of 24 comparison simulations comparing a cold dark matter (CDM) and 3.5 keV warm dark matter (WDM) particle, primarily using a dust-to-metals (DTM) ratio to post-process the dust in the simulation, with alternative methods explored. We find a preference for higher dust masses in CDM over WDM, with a median increase of 6.6 per cent in the CDM sample at redshift $z=0$, and with a peak median difference of 25 per cent at $z\approx0.25$. Both CDM and WDM samples are highly concentrated, with approximately 90 per cent of the dust located in the inner region of the halo, though slightly higher in CDM. Contrary to previous work, we find that the Gini coefficient does not consistently show a difference in comparisons of dark matter models, with differences primarily driven by high sensitivity to the unique structure of individual haloes. Finally, we conduct an analysis of some common dust scaling relations, finding that the differences between the two samples in the dust-to-stellar mass and dust-to-SFR functions do not constitute a statistically significant difference at  $z=0$ with any dust production mechanism. However, at redshifts where the difference in dust mass is greatest, the alternative dust production methods show larger separations between CDM and WDM samples in total mass and scaling relations.
\end{abstract}




\section{Introduction}\label{sec:Intro}
The mass budget of the Universe is primarily dominated by some minimally interacting form of matter, or dark matter. The current standard model of dark matter is so-called cold dark matter (CDM), where the `cold' term refers to any particular particle model that has a non-relativistic thermal velocity at the time of decoupling in the very early Universe. The CDM model has proved highly successful at recreating observed statistics on large scales, from the angular power spectrum of the Cosmic Microwave Background \citep{aghanim_planck_2020a}, to large-scale clustering of galaxies and the presence of baryonic acoustic oscillations \citep{eisenstein_detection_2005, abbott_dark_2018}, down to the density profiles of individual dark matter haloes \citep{navarro_structure_1996}.

Despite the success of the CDM model at capturing large-scale astronomical observations, there are some common historical and ongoing tensions between cosmological simulations of a CDM universe and small-scale galaxy observations \citep{bullock_smallscale_2017}. Some of the more well-known historical tensions include the core-cusp problem \citep{moore_evidence_1994, swaters_central_2003, deblok_corecusp_2010}, the too-big-to-fail problem \citep{boylan-kolchin_too_2011}, and the missing satellites problem \citep{moore_dark_1999, klypin_galaxies_1999}. These historical tensions are often addressed through various implementations of baryonic physics in simulations, but tensions such as the diversity of rotation curves in dwarf galaxies \citep{oman_unexpected_2015} and the apparent plane of satellites observed around the Milky Way \citep{kroupa_great_2005, metz_spatial_2007, buck_simulated_2016, pawlowski_planes_2018} still present a challenge for simulations to match to observations \citep[see][for further discussion]{sales_baryonic_2022}. 

In light of the tensions between CDM and observations, as well as the lack of confirmed, robust particle detection \citep{aprile_dark_2018,aprile_first_2023, adhikari_strong_2021, crocker_gammaray_2022, coarasa_anais112_2024, james_dama_2024}, a wide variety of alternative dark matter candidates have been proposed. Some of these alternatives include warm dark matter (WDM) \citep{bode_halo_2001, viel_warm_2013, lovell_general_2020}, self-interacting dark matter \citep{carlson_selfinteracting_1992, atrio-barandela_interactinghotdark_1997, spergel_observational_2000}, dark matter interacting with standard model particles \citep{boehm_constraining_2001, boehm_constraints_2005}, and fuzzy dark matter \citep{irsic_first_2017, dentler_fuzzy_2022}. In particular, WDM has long been posed as a compelling alternative to CDM, where `warm' describes a class of particles with relativistic velocities at the moment of decoupling. Their velocity induces a free-streaming scale below which matter overdensities are erased. In practice, this means that the WDM power spectrum is exponentially suppressed in the early Universe below the free-streaming scale, which leads to a different pattern of structure formation compared to CDM. WDM models are often used to relax the small-scale assumptions of a CDM model, especially in relation to tensions such as the missing satellites problem, where the inherent reduction in low-mass haloes reduces the number of expected Milky Way satellite galaxies.

However, similar solutions to many of the small-scale tensions between observations can be resolved in cold dark matter hydrodynamical simulations through various feedback mechanisms. Both high levels of supernova-driven feedback \citep[e.g.][]{sawala_apostle_2016}, and efficient gas heating during the epoch of reionisation \citep[e.g.][]{bullock_reionization_2000, benson_effects_2002} can be used to solve the tensions such as the missing satellites problem \citep[see][for reviews]{bullock_smallscale_2017, sales_baryonic_2022}. Solving these tensions with galaxy formation models creates an interesting area of degeneracy. The differences between dark matter models are most likely to manifest at lower mass scales, but this is also the region where differences in galaxy formation models have the greatest effect. In \cite{ussing_using_2024}, hereafter referred to as U24, a novel method for breaking this degeneracy was explored, using dust properties of simulated galaxies to show differences between CDM and WDM models in a single set of comparison simulations. 

Dust provides a compelling tool in distinguishing between different dark matter models, specifically because metrics as fundamental as the total mass and dust scaling relations are dependent on the implemented dust model \citep[e.g.][]{popping_dust_2017, aoyama_cosmological_2018, triani_origin_2020, granato_dust_2021, vijayan_first_2022, dubois_galaxies_2024}. Dust is the byproduct of several astronomical processes, mainly supernova \citep{aguirre_intergalactic_1999, bianchi_dust_2007} and stellar winds from asymptotic giant branch stars \citep{valiante_stellar_2009}, and as such is sensitive to changes in the supernova rate. Increasing or decreasing the rate at which supernovae occur has a significant impact on the rate at which low-mass galaxies can form, hence their use in tuning hydrodynamical simulations to alleviate the missing satellites problem, but this also feeds into the dust properties.

In U24, it was found that tuning the supernova efficiency in cold and warm dark matter simulations created similar satellite galaxy populations with different dust properties. The CDM simulation had a greater overall dust mass than the WDM simulation, with a greater percentage of the total dust mass found in the central 15kpc of the halo, which was used as a rough approximation of the central galaxy disc. The dust between 15kpc and the halo radius was also more concentrated in its distribution, as given by the Gini score for that region. The Gini score is generally used to quantify the inequality of data across bins in a sample, which has previously been used to classify galaxy morphology \citep[e.g.][]{lotz_new_2004}. The Gini score here was adapted to measure the concentration of dust surface density in the region between 15kpc and \rthc{}, as this provides a straightforward characterisation of the dust distribution. The U24 results found a significant difference in the total dust mass, where the fiducial CDM simulation produced $\sim4\times$ more dust than the fiducial WDM simulation, and both CDM simulations also had a systematically higher Gini score than the WDM simulations. U24 showed that dust, as an integrated property of galaxies in the simulations, can break the degeneracy between different galaxy formation and dark matter models, but is also sensitive to the halo assembly history. Critically, the U24 analysis was based on a single set of comparison simulations, so in this work, we extend the suite to 24 individual comparison simulations to determine the significance of the dust differences in a larger sample and characterise the variance.

This paper is structured as follows: the codes and processes are described in Section \ref{sec:Methods}, and we present initial results in Section \ref{sec:Results}. We then discuss these results and their comparisons to previous work in Section \ref{sec:Disc} and conclude in Section \ref{sec:Conc}. For all simulations we follow a flat Planck18 cosmology \citep{aghanim_planck_2020}, where $H_0= 100h$ km/s/Mpc and $h = 0.6688$, $\Omega_m=0.321$, $\Omega_{\Lambda}=0.679$, $\Omega_{b}=0.045$, $\sigma_8$=0.840, and $n_s=0.9626$.

\section{Methods}\label{sec:Methods}
\subsection{Simulation setup}\label{sec:methods_sim_setup}
Following the processes outlined in U24, we create a suite of hydrodynamical zoom-in simulations of Milky Way-like galaxies to compare the two different dark matter scenarios, cold and warm dark matter, using the \gadget{} simulation code \citep{springel_simulating_2021}. We first identify Milky Way-like galaxies in low-resolution CDM dark matter only simulations, with $512^3$ particles in a $65^3$\hMpc{}, using a halo mass criterion to select Milky Way-like targets. It is sufficient to run only one parent simulation, as the resolution is insufficient to distinguish between the cold and warm dark matter models in this setup. We make two extensions to the methodology of U24 target selection criteria, in terms of the halo mass and exclusion radius. Firstly we bring the target halo mass selection criterion in line with observations from \cite{bland-hawthorn_galaxy_2016}, with an \mthc{} between $0.55-2.62\times10^{12}$ \msun{}, rather than $ 1- 2\times10^{12}$ \msun{} used previously, where \mthc{} is the mass enclosed in a sphere of radius \rthc{} such that the enclosed average density $\rho$ is 200$\times$ the critical density of the universe, $\rho_c$. Secondly, we relax the isolation criteria slightly, so that no interfering haloes (haloes whose mass is greater than half of the allowed Milky Way mass range) are present within 7\hMpc{}, rather than 8\hMpc{}. This allows us to select more resimulation targets from a single parent simulation.

After target selection, we resimulate each target with a cold and 3.5 keV warm dark matter model in a zoom-in hydrodynamical simulation, using the same random seed for each set of simulations. A particle mass of 3.5 keV is in tension with several recent published constraints \citep{viel_warm_2013, nadler_constraints_2021, dekker_warm_2022}, but following the U24 methodology we employ a lighter mass to try to ensure that differences between small scale structure arise due to the different dark matter model, without straying too far from observed lower mass limits \citep[e.g.][]{benito_implications_2020, newton_constraints_2021}.



We use the \cite{viel_constraining_2005} transfer function to generate the input power spectra, which also gives us WDM half-mode mass, $M_{hm}\simeq2.3 \times10^8$\msun{}, which is the mass below which we expect significant divergence between the CDM and WDM halo mass functions. We run two simulations for each dark matter model with different supernova strengths, using the included \cite{springel_cosmological_2003} baryonic model, with both a 5\% and 10\% supernova efficiency (defined as the $\beta$ parameter in this model), creating four total simulations for each Milky Way analogue. Again, we nominate the CDM simulation with 10\% supernova feedback efficiency and the WDM with 5\% efficiency as our fiducial models. For succinctness, the results presented below show only the results for the fiducial models.

\subsection{Dust properties}\label{sec:methods_dust_prod}

All simulations are subsequently processed with \powderday{} \citep{narayanan_powderday_2021}, which is built on \textsc{yt} \citep{turk_yt_2011}, \textsc{hyperion} \citep{robitaille_hyperion_2011}, and \textsc{fsps} \citep{conroy_propagation_2010}, to produce the dust statistics from redshift $z=4$ to $z=0$. \powderday{} includes several dust prescriptions for assigning dust masses to gas particles, including a \dtm{} ratio, the dust-to-gas ratio from \cite{remy-ruyer_gastodust_2014}, and a `best-fit' combination dust-to-metals/gas ratio from the machine learning model from \cite{li_dusttogas_2019}. We compare the results of all production methods. For succinctness, we primarily present the results using a \dtm{} ratio of 0.4 \citep{dwek_evolution_1998}, knowing that a constant \dtm{} prescription is a simplification of more complex astrophysical systems \citep[e.g.][]{devis_systematic_2019}, in this case only tracking metal content of gas particles, but makes for a more straightforward comparison, with a discussion of the other production methods in relevant sections. We compare the simulated dust properties, despite functioning as a rescaled metallicity, in order to compare with observed dust properties, which, as previously mentioned, are sensitive to supernova rates, the metric that we use to tune the CDM and WDM simulations. Further results for the dust-to-gas and best-fit methods are shown in Appendix \ref{app:appendix_A}. 

Following U24 we use the Gini coefficient, or score, to quantify the dust in the outer regions of the halo. The Gini coefficient, G, follows
\begin{equation}\label{eq:gini}
    G = \frac{\sum\limits_{i=1}^n \sum\limits_{j=1}^n |x_i - x_j |}{2n^2\overline{x}},
\end{equation}
where $x_{i}$ is the dust mass in the $i$-th bin. We use $2.5h^{-1}$ckpc bins in a 2D projection to discretise the dust mass between redshifts $z=4$ and $z=0$. To avoid projection effects, we rotate the 2D projections 25 times through 360$^{\circ}$ each around the $x$, $y$, and $z$ axes, for 75 total realisations. Additionally, the central galaxies inside the halo tend to saturate the Gini score, and so, as in U24, we mask the central regions, although we expand the centrally masked region to a radius of 25 kpc rather than 15 kpc, to account for larger galactic disks. The amount of dust beyond \rthc{} is negligible and of little use in either analysis or observations, and so these are also masked. In total, we created 24 sets of comparison simulations, including the simulations used in U24. 

\section{Results}\label{sec:Results}
\subsection{Satellite mass functions}\label{sec:results_satellites}

\begin{figure}
    \begin{subfigure}[]{\columnwidth}
    \centering
        \includegraphics[width=\textwidth]{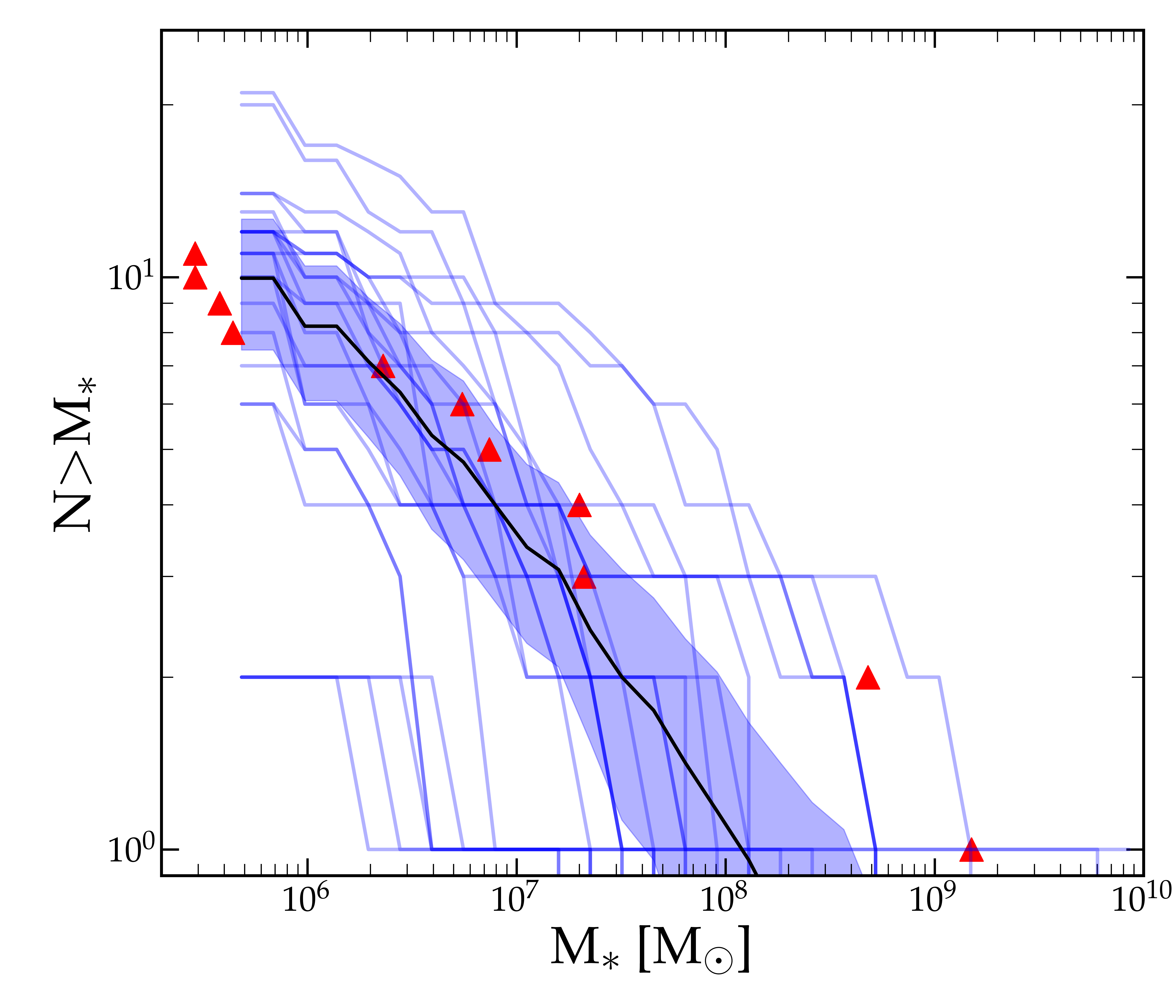}
        \caption{CDM 10\%}
        \label{fig:satellite_pop_cdm_sn_010}
    \end{subfigure}
    \hfill
    \begin{subfigure}[]{\columnwidth}
    \centering
        \includegraphics[width=\textwidth]{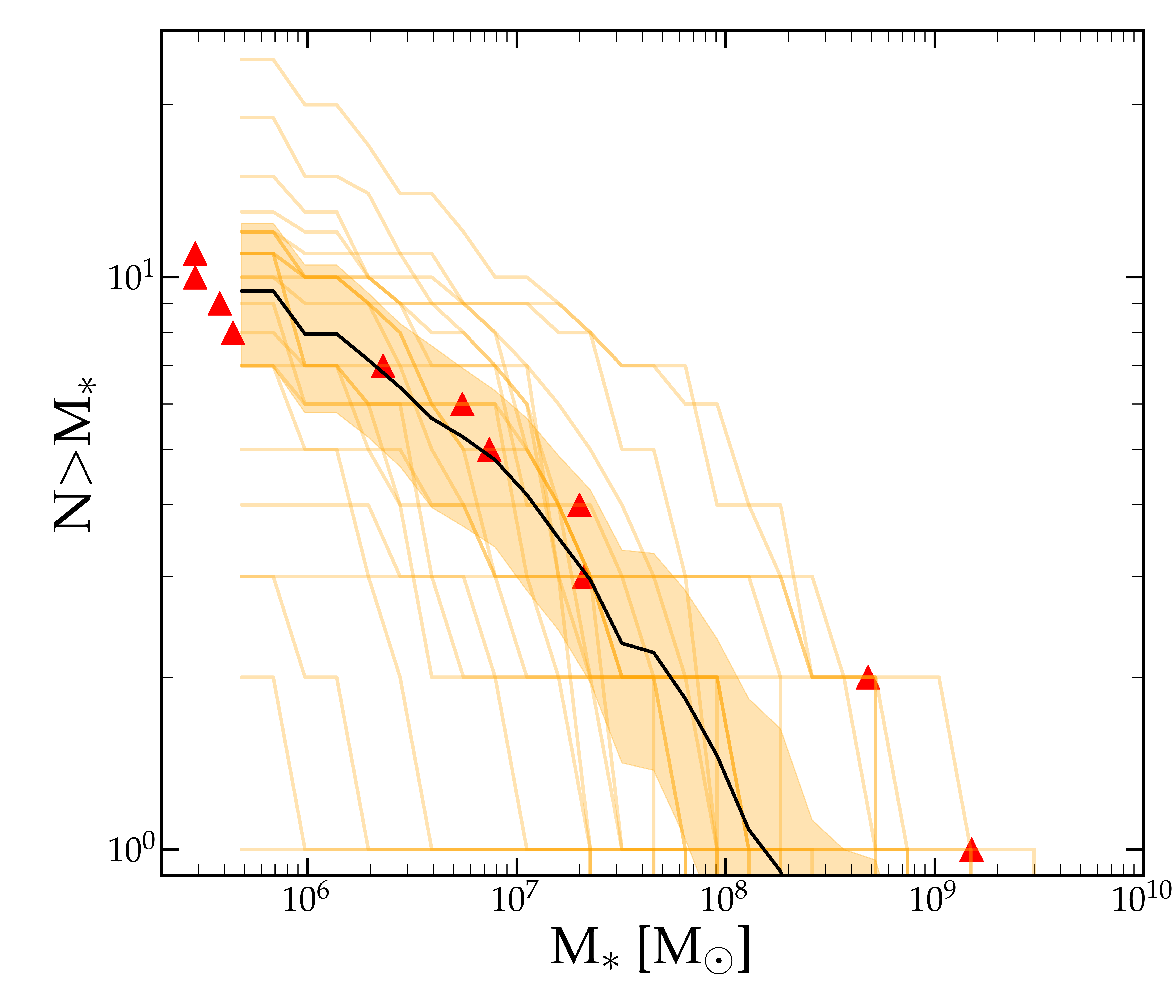}
        \caption{WDM 5\%}
        \label{fig:satellite_pop_wdm_sn_005}
    \end{subfigure}
    \hfill
    \caption{The ensemble satellite mass function for both cold and warm dark matter simulations. The black line in each plot shows the mean number of satellites at each mass, with the shaded region showing 99\% confidence intervals calculated using a bootstrap error estimation. The red triangles show the mass of the 11 most luminous satellite galaxies around the Milky Way, with values taken from \protect\cite{lovell_matching_2023}.}
    \label{fig:satellite_pop_all}
\end{figure}

We begin the analysis with the satellite mass function for all simulations, which are shown in \ref{fig:satellite_pop_all}. We show the fiducial models, a cold dark matter simulation with 10\% supernova efficiency and a warm dark matter simulation with a 5\% efficiency, noting that we have rescaled the mass values to factor out $h$ from the units. For this work, we define a satellite galaxy as a subhalo with at least a singular star particle and at least 100 total particles. We use the 100 particle limit to minimise the effects of artificial fragmentation known to inflate small halo numbers in WDM simulations \citep{power_spurious_2016, mostoghiupaun_tidal_2025}. Spurious haloes in these WDM simulations dominate the halo mass function below $M_{lim} \simeq 2\times 10^8$\msun{} \citep{wang_discreteness_2007, lovell_properties_2014}, which is below the mass of a halo with 100 particles. The mass functions are binned into logarithmically spaced 30 bins between $4\times10^5$ and $1\times10^{10}$\msun{}, and we then calculate the mean number of satellites in each bin, with a bootstrapped error showing a 99\% confidence region. 

We see good agreement between the two models and the observed satellites, with the cold dark matter simulations producing slightly more satellites on average. However, the chance of randomly selecting a Milky Way-sized halo with subhaloes that have masses matching both Magellanic clouds was quite low, as they both fall outside the 99\% confidence interval of the mean in this suite of simulations. This is also found observationally, where \cite{robotham_galaxy_2012} found that Milky Way-Magellanic Cloud systems are very rare in the GAMA survey, and generally zoom-in simulations preselect haloes that contain a subhalo of similar size to the Large Magellanic Cloud \citep[e.g.][]{buch_milky_2024}. The goal of this work, however, was to create degenerate simulations, rather than exactly recreating the Milky Way.

\subsection{Dust masses across the suite}\label{sec:results_dust_mass}

We see general agreement with the trend initially identified in U24 when looking at the total dust masses for all simulations using the \dtm{} production method. This can be seen in the masses in \ref{tab:suite_dtm_dust_masses}, which show that the fiducial cold dark matter simulations tend to produce more dust than warm dark matter simulations, with a median (mean) CDM to WDM dust mass ratio of 1.07 (1.31) $\pm$ 0.85. The majority of simulations do show at least a minor increase in the dust mass in the cold dark matter simulations, although seven out of the 24 total comparison simulations have a greater dust mass in the warm dark matter simulations. The median and mean dust masses show moderate to significant increases with changes in the dust production method, shown in \ref{tab:suite_rr_dust_masses} and \ref{tab:suite_li_bf_dust_masses}, which is discussed further in \ref{sec:discussion_dust}.

\begin{table}
    \centering
    \begin{tabular}{c|c|c|c}
        \hline
        Simulation  & CDM 10\%  & WDM 5\% & CDM/WDM\\
        ID          & ($\times10^7$\msun{}) & ($\times10^7$\msun{}) &  mass ratio\\
        \hline
        ID-00 & 12.34 & 7.49   & 1.65  \\
        ID-01 & 10.35 & 8.12   & 1.28  \\
        ID-02 & 12.20 & 11.24  & 1.09  \\
        ID-03 & 18.33 & 14.28  & 1.28  \\
        ID-04 & 1.61  & 1.73   & 0.93  \\
        ID-05 & 7.19  & 2.04   & 3.52  \\
        ID-06 & 11.52 & 17.32  & 0.67  \\
        ID-07 & 3.78  & 3.74   & 1.01  \\
        ID-08 & 3.35  & 2.43   & 1.38  \\
        ID-09 & 5.57  & 4.25   & 1.31  \\
        ID-10 & 4.81  & 5.67   & 0.85  \\
        ID-11 & 5.60  & 6.14   & 0.91  \\
        ID-12 & 2.06  & 4.78   & 0.43  \\
        ID-13 & 4.14  & 4.11   & 1.01  \\
        ID-14 & 4.05  & 4.21   & 0.96  \\
        ID-17 & 6.00  & 4.98   & 1.20  \\
        ID-28 & 1.38  & 1.05   & 1.31  \\
        ID-34 & 1.40  & 3.68   & 0.38  \\
        \textbf{ID-46} & 4.56  & 1.05   & 4.35  \\
        ID-57 & 6.27  & 5.99   & 1.05  \\
        ID-61 & 6.12  & 4.25   & 1.44  \\
        ID-70 & 8.39  & 6.48   & 1.30  \\
        ID-80 & 9.18  & 9.05   & 1.01  \\
        ID-93 & 9.35  & 9.10   & 1.03  \\
        \hline
        Median & 5.80 & 4.88 & 1.07  \\
        Mean   & 6.65 & 5.97 & 1.31  \\
        $\sigma$ & 4.08 &	3.94 &	0.85 \\
        \hline
    \end{tabular}
    \caption{The dust masses for each simulation within \protect\rthc{} using the \protect\dtm{} method, with summary statistics in the bottom rows. 17 of the 24 simulations produce more dust in the cold dark matter simulations than the warm dark matter counterparts, seven simulations produce more dust in the fiducial warm dark matter simulations, though several simulations have quite low mass ratios. The bolded simulation is the one used in U24.}
    \label{tab:suite_dtm_dust_masses}
\end{table}

\begin{figure}
    \centering
    \includegraphics[width=\columnwidth]{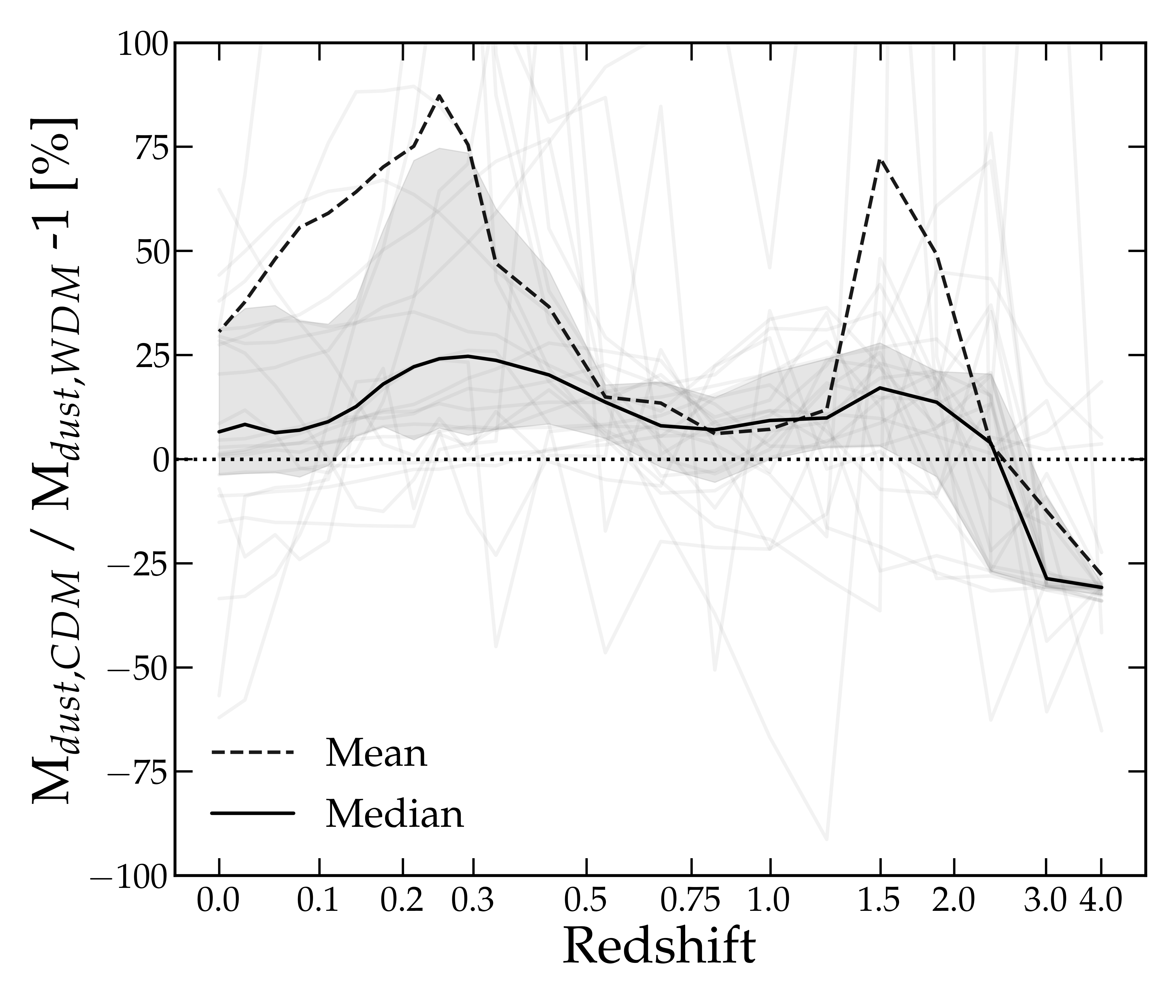}
    \caption{The redshift evolution of the ratio of the CDM and WDM dust mass ratio as a percentage difference of the dust masses in each simulation. The light grey lines show the ratios for each simulation, and the solid and dashed lines show the median and mean ratios, respectively. The grey shaded region shows the bootstrapped 99 per cent confidence interval on the median.}
    \label{fig:dust_mass_ratio_evo}
\end{figure}

To explore the evolution of the dust mass as a function of redshift, we show the evolution of the differences in dust mass in \ref{fig:dust_mass_ratio_evo}. Each light grey line shows the evolution of the total dust mass comparison for a given simulation, while the black solid line and dashed lines show the median and mean difference, respectively, for the sample. The grey shaded region shows the 99 per cent confidence interval using a bootstrap resampling method. We find that the median dust mass ratio consistently favours the CDM suite after $z=2$, where the increase in dust mass generally falls between 5\% and 25\%, peaking around $z=0.25$. As redshift decreases, the median difference across the suite of simulations decreases, approaching parity. 

By $z=0$, the bootstrapped confidence interval on the median difference has a large spread, which includes a zero per cent difference. However, between redshifts $z\sim1.5-1$ and $z\sim0.5-0.15$, where the difference is more pronounced, this confidence interval is exclusive of a zero per cent change, indicating these redshift ranges are promising points to investigate that show differences in observations. The increase in dust mass during the earlier epoch is driven by increased metallicity from higher star formation rates (SFR) earlier in the CDM sample. This increased metallicity then falls as the SFR returns to parity, though remaining higher up until $z=0$. While the consistent increase in dust mass in the CDM suite is promising, the overall stochasticity of the individual comparisons, and the loss of a clear signal at later redshifts, remain barriers to the use of dust as a tool in dark matter research.

\subsection{Structural properties of the dust}\label{sec:dust_structure}
\subsubsection{Dust concentrations}\label{sec:dust_structure_percentages}
Following the U24 analysis, we now study the structural properties of the dust. All haloes typically have quite high dust concentrations, most having 85-95\% of the dust located within the inner 25 kpc, listed in \ref{tab:suite_dtm_outer_dust_masses}. As mentioned earlier, the definition of the central region in this study is broadened from 15 kpc, as used in U24, to 25 kpc to account for a wider variety of galactic discs in the larger sample. The average dust mass in the inner region is higher in the cold dark matter simulations, at an average of 89.1\% compared to 85.7\% in the warm dark matter haloes. Consequently, we see a slightly higher mean relative dust mass, and though not shown here, a higher mean absolute dust mass, in the outer regions of the warm dark matter haloes. 

Simulation ID-02 is a standout case in the sample, where most of the dust is located outside the central galaxy, which is caused by a relatively gas-poor central galaxy and a very gas-rich infalling satellite galaxy. This infalling satellite drives high total dust masses and singles the halo out as the only one with less than 50\% in the central region; however, there is no significant difference in the dust mass between the dark matter models. Simulation ID-46, used in U24, is a standout simulation in terms of dust mass ratio, at approximately 4.5. The reasons this simulation stands out so much are discussed in \ref{sec:discussion_pap_1}. Overall, we find only a slight preference for a higher dust mass concentration in the CDM sample over the WDM sample. 

\begin{table*}
    \centering
    \begin{tabular}{c|c|c|c|c}
        \hline
        Simulation & \multicolumn{2}{c}{CDM 10\%} &\multicolumn{2}{c}{WDM 5\%} \\
        ID  & \% ($r\leq25\text{kpc}$) &  \% ($25\text{kpc} < r \leq R_{200\text{c}})$   & \% ($r\leq25\text{kpc}$) &  \% ($25\text{kpc} < r \leq R_{200\text{c}})$   \\
        \hline
        ID-00 & 96.0\% & 4.0\%. & 91.3\% & 8.7\%   \\
        ID-01 & 98.2\% & 1.8\%  & 97.2\% & 2.8\%   \\
        ID-02 & 18.5\% & 81.5\% & 9.6\%  & 90.4\%  \\
        ID-03 & 91.7\% & 8.3\%  & 87.5\% & 12.5\%  \\
        ID-04 & 85.9\% & 14.1\% & 85.3\% & 14.7\%  \\
        ID-05 & 92.8\% & 7.2\%  & 69.6\% & 30.4\%  \\
        ID-06 & 92.3\% & 7.7\%  & 94.8\% & 5.2\%   \\
        ID-07 & 96.5\% & 3.5\%  & 90.6\% & 9.4\%   \\
        ID-08 & 97.1\% & 2.9\%  & 95.6\% & 4.4\%   \\
        ID-09 & 96.0\% & 4.0\%  & 93.2\% & 6.8\%   \\
        ID-10 & 71.6\% & 28.4\% & 67.0\% & 33.0\%  \\
        ID-11 & 89.7\% & 10.3\% & 88.3\% & 11.7\%  \\
        ID-12 & 84.9\% & 15.1\% & 83.4\% & 16.6\%  \\
        ID-13 & 95.9\% & 4.2\%  & 92.8\% & 7.2\%   \\
        ID-14 & 95.2\% & 4.8\%  & 93.3\% & 6.7\%   \\
        ID-17 & 94.1\% & 5.9\%  & 91.2\% & 8.8\%   \\
        ID-28 & 93.9\% & 6.1\%  & 90.4\% & 9.6\%   \\
        ID-34 & 83.9\% & 16.1\% & 92.3\% & 7.7\%   \\
        \textbf{ID-46} & 87.8\% & 12.3\% & 77.6\% & 22.4\%  \\
        ID-57 & 95.7\% & 4.3\%  & 93.3\% & 6.7\%   \\
        ID-61 & 96.9\% & 3.1\%  & 95.2\% & 4.8\%   \\
        ID-70 & 99.3\% & 0.7\%  & 99.3\% & 0.7\%   \\
        ID-80 & 97.4\% & 2.6\%  & 94.4\% & 5.6\%   \\
        ID-93 & 87.5\% & 12.5\% & 82.3\% & 17.7\%  \\
        \hline
        Median & 94.0\% & 6.0\%  & 91.3\% & 8.8\%  \\
        Mean   & 89.1\% & 10.9\% & 85.7\% & 14.3\% \\ 
        \hline
    \end{tabular}
    \caption{The percentage of how much dust lies within or beyond the inner 25 kpc, within \rthc{}. Most simulations produce highly concentrated dust distributions and show a slightly higher relative dust mass in the inner region of the CDM haloes compared to the WDM haloes, which have a slightly higher mean percentage of the dust mass outside the inner 25 kpc. As above, the bolded simulation is the simulation presented in U24.}
    \label{tab:suite_dtm_outer_dust_masses}
\end{table*}

\subsubsection{Gini trends across the suite}\label{sec:results_gini_trend}

\begin{figure}
    \centering
        \includegraphics[width=\columnwidth]{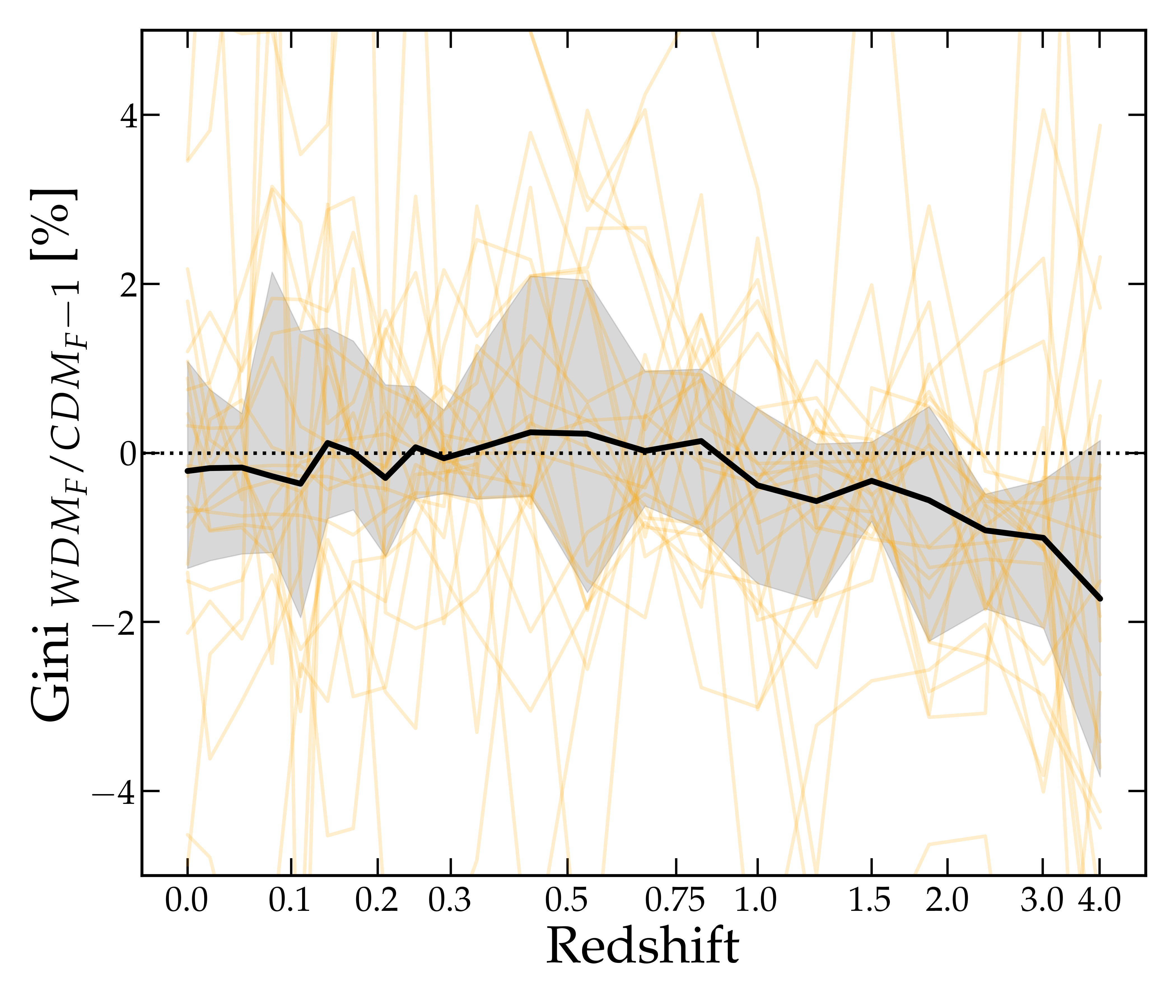}
    \caption{The difference between the Gini scores in the fiducial WDM and CDM simulations. The median and 99\% confidence interval are shown as the black line and grey shaded region, respectively. Despite a large spread in differences, the median for the majority of the evolution is consistent with zero difference. }
    \label{fig:gini_difference_all}
\end{figure}

Previous work in U24 demonstrated that the Gini coefficient can be used to illustrate differences in the dust properties of cold and warm dark matter haloes. This is examined in greater detail here across the full suite of simulations. The evolution of the ratio of Gini scores between cold and warm dark matter suites is shown in \ref{fig:gini_difference_all}. This figure is slightly updated from Figure 5 in U24; the redshift range is increased to $z=4$, and the shaded region shows the 99\% confidence interval on the median, rather than a 68\% confidence interval. The combined mean difference generally hovers around zero, indicating very little difference between dust concentrations in cold and warm dark matter scenarios in a larger sample. There is a slight indication that WDM simulations show lower concentrations at $z>1.5$, which may be due to a physical delay in assembly or to a numerical artifact in the stellar modelling at higher redshifts. Regardless, there is still a large scatter in the individual scores, and the confidence interval is still close to zero difference, hence more testing would be required to verify this behaviour. Overall, the differences between simulations are very stochastic, with little identifiable trend from one simulation to the next. 

In Appendix \ref{app:appendix_B}, we explore the ratio of the Gini scores between the two cold dark matter simulations, shown in \ref{fig:gini_difference_dtm_cdm_sn_005}. The difference in scores between the two CDM samples still shows a reasonable amount of scatter, largely ranging between -4 to 4 per cent, but to a lesser extent than the difference between the two fiducial simulations. The median difference between them also indicates that the lower supernova efficiency drives slightly lower average Gini scores at all redshifts, although still with errors inclusive of zero. The smaller scatter and slight reduction in dust concentration appears to suggest that a change in the dark matter model can drive moderate stochastic differences in individual simulations, but the lower supernova efficiency drives a slight change for the suite as a whole.

\begin{figure}
    \begin{subfigure}{\columnwidth}
    \centering
        \includegraphics[width=\textwidth]{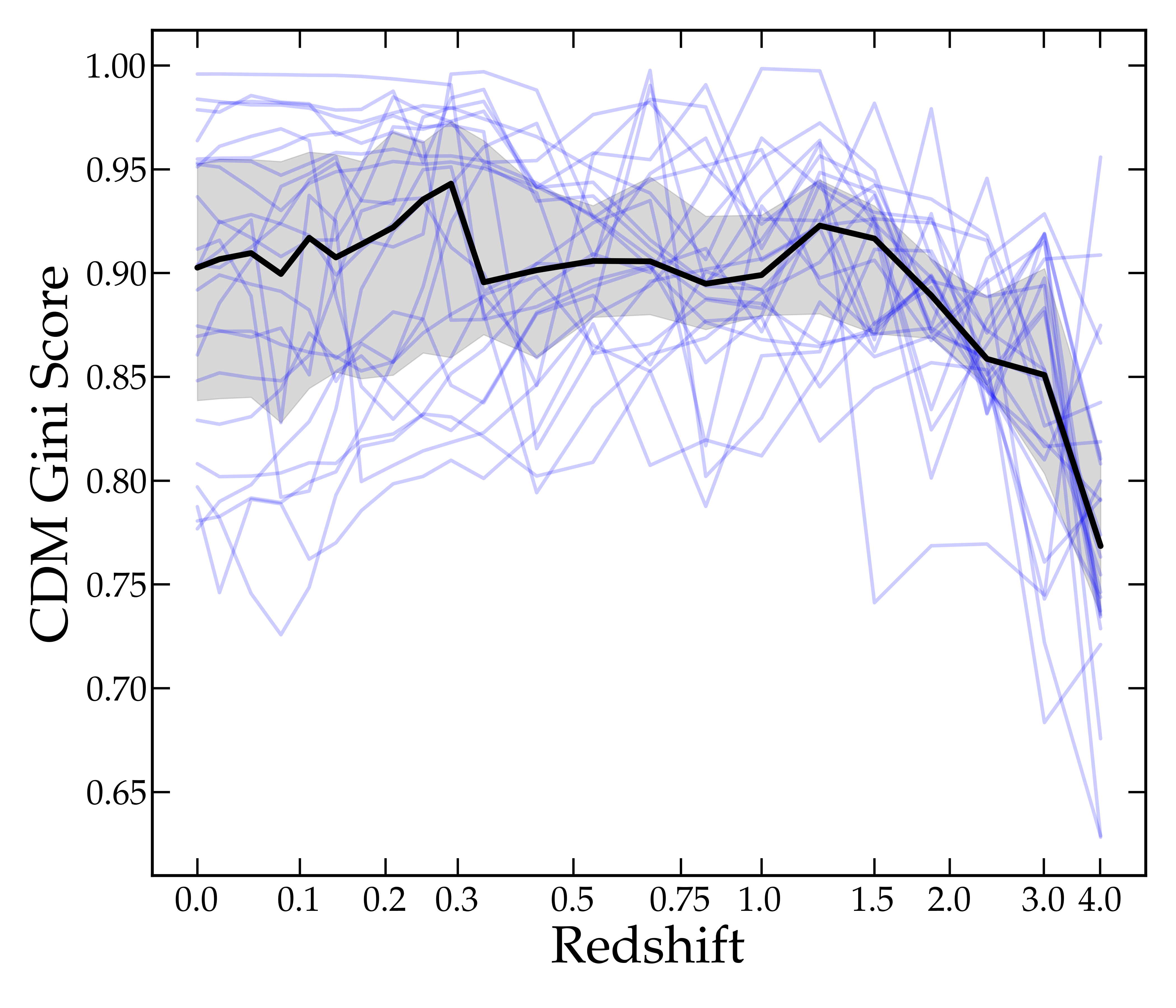}
        \caption{CDM 10\%}
        \label{fig:gini_score_dtm_cdm_sn_010}
    \end{subfigure}
    \hfill
    \begin{subfigure}\columnwidth
    \centering
        \includegraphics[width=\textwidth]{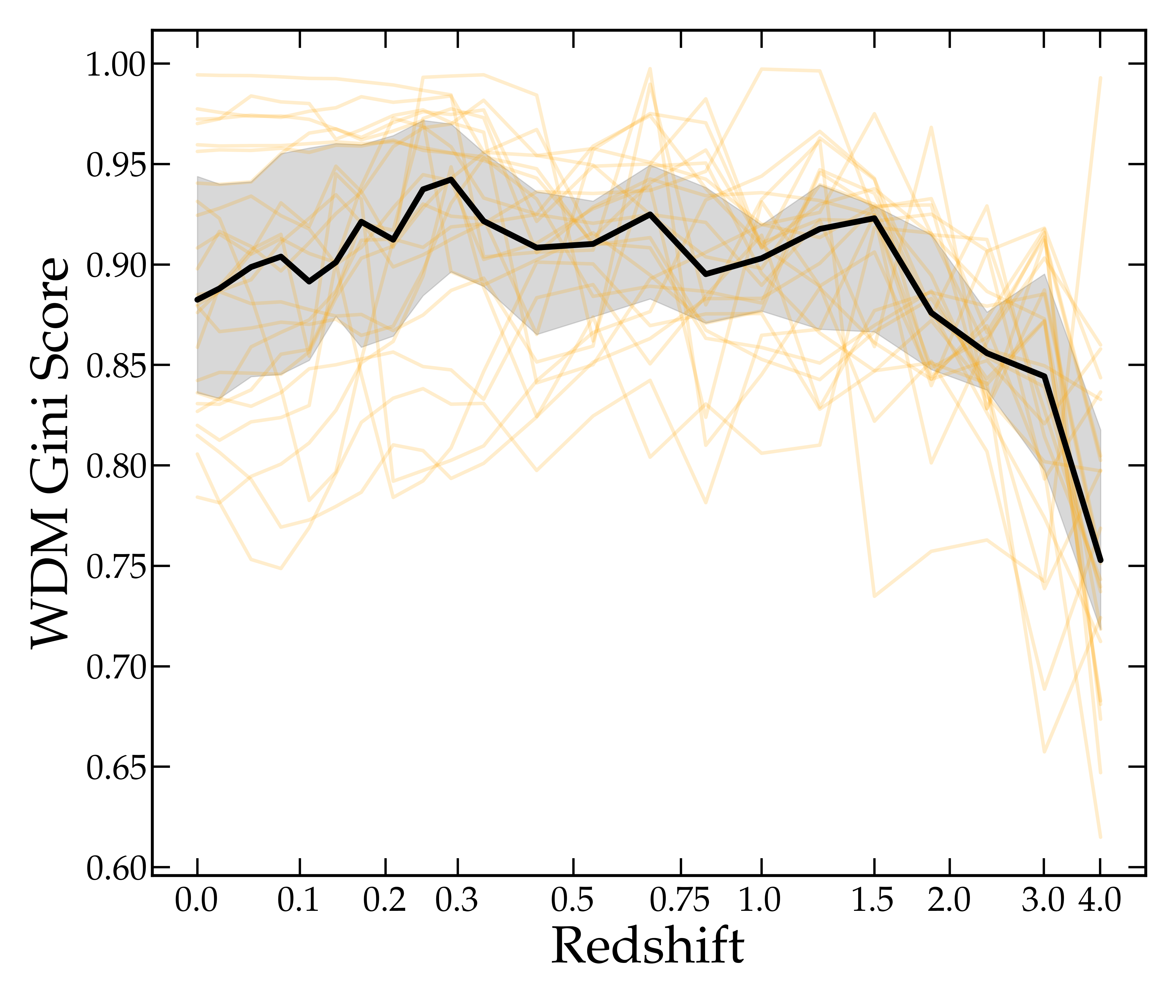}
        \caption{WDM 5\%}
        \label{fig:gini_score_dtm_wdm_sn_005}
    \end{subfigure}

    \caption{\textit{Top:} The individual Gini scores for every simulation for the fiducial CDM sample. \textit{Bottom:} The Gini scores for the fiducial WDM sample. As above, the black line and grey region show the sample median and 99\% confidence interval in both top and bottom plots. }
    \label{fig:gini_score_all}
\end{figure}

There is some interesting evolution in the absolute Gini scores, shown in \ref{fig:gini_score_all}, that is common to all simulations. There is a large spread in the scores, but they all show significant increases in concentration between $z\simeq4$ to $z\simeq1$ as the mass build-up is quite rapid. Beyond $z=4$, the time resolution of the snapshots is insufficient to properly track the evolution of the haloes, and so these are omitted from the analysis. Once this rapid mass consolidation phase ends, around $z\simeq1$, the Gini scores diverge rapidly, but the mean scores trend slightly downward. This trend may be caused by the progression of marginally more concentrated regions merging into the central galaxy and falling under the masked area. However, the large scatter in the Gini scores, ranging from 0.7 to 1 at $z=0$, underscores that any trend is dominated by numerical variance, at least within this suite of simulations. 

\subsection{Differences in common scaling relations}

As a further check on how the different dust mass samples propagate into observable metrics, we compare some common dust scaling relations. We combine this process with some preliminary statistical tests to determine the likelihood that these samples represent unique distributions, noting that this analysis is intended to be indicative of a future comparison to observations, given the simplistic modelling, and that the sample is still relatively small. 

\begin{figure}
    \centering
    \includegraphics[width=\columnwidth]{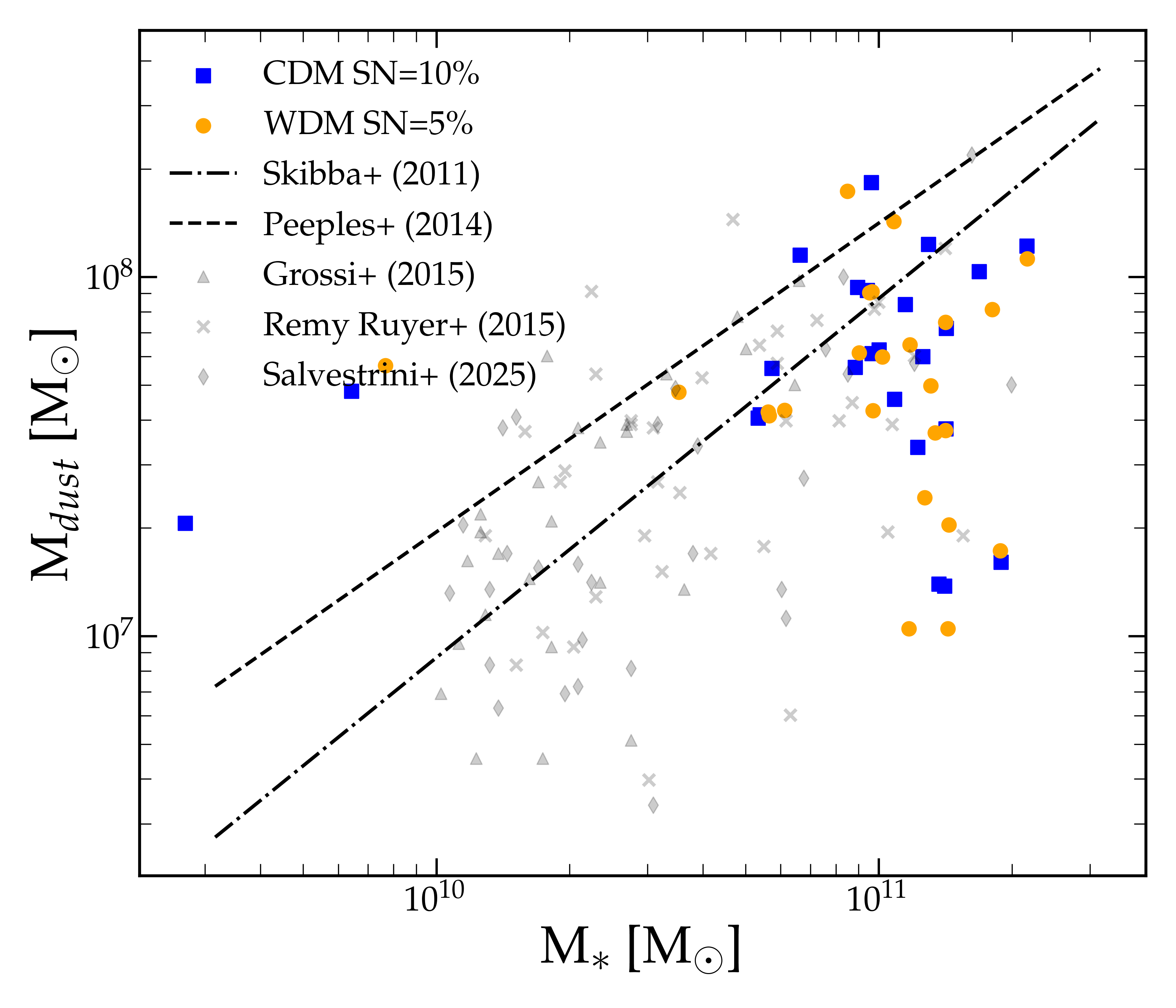}
    \caption{The total dust mass in each halo as a function of the total stellar mass, where the blue squares show the CDM sample and the orange circles show the WDM sample. The two black lines show observational fits to the dust-to-stellar mass function; the dash-dot line shows the fit from \protect\cite{skibba_emission_2011}, and the dashed line shows the fit from \protect\cite{peeples_budget_2014}. Grey triangles, crosses, and diamonds show a subset of observational samples from \protect\cite{grossi_herschel_2015}, \protect\cite{remy-ruyer_linking_2015}, and \protect\cite{salvestrini_exploring_2025}, respectively.}
    \label{fig:dust-to-stellar-mass}
\end{figure}

In \ref{fig:dust-to-stellar-mass} we show the dust mass as a function of the stellar mass for each simulation in the fiducial CDM and WDM suites, as well as a subset of observational data from \cite{grossi_herschel_2015}, \cite{remy-ruyer_linking_2015}, and \cite{salvestrini_exploring_2025}, and fits from \cite{skibba_emission_2011} and \cite{peeples_budget_2014}. The subsampling was to limit the data to a range relevant to our simulations; the full datasets include galaxies with much lower stellar masses.  While both CDM and WDM samples are slightly below the observed fits, they fall well within the scatter of observations at these stellar masses, though neither sample appears to match observations better than the other. To determine if the samples themselves are statistically unique, we employ a 2-dimensional version of the Kolmogorov-Smirnov (KS) test, known as the Fasano-Franceschini (FF) test \citep{fasano_multidimensional_1987}, using the Python implementation described in \cite{chow_decametersized_2025}\footnote{Available at \url{https://github.com/wmpg/fasano-franceschini-test.git}}. The $p$-value of this test, or a measure of the likelihood that two samples are drawn from the same underlying distribution, has a value of 0.9, far greater than the commonly used threshold value of 0.05 to reject the null hypothesis, indicating that we cannot conclude that the distributions are different. Additionally, the scatter in the observations spans both samples, making inference difficult from this sample alone.

\begin{figure}
    \centering
    \includegraphics[width=\columnwidth]{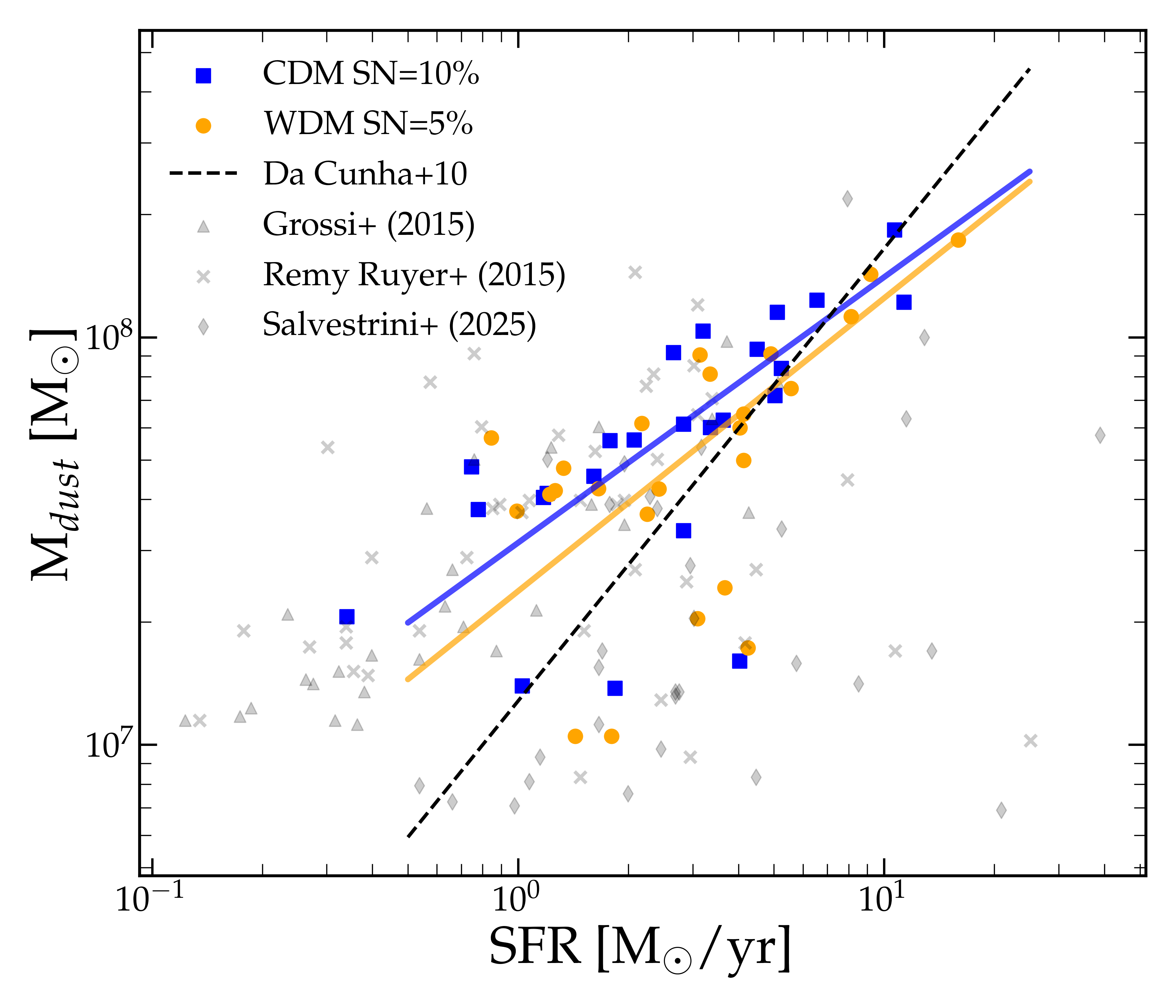}
    \caption{The total dust mass in each simulation as a function of the SFR. Again, the blue squares show the CDM sample and the orange circles show the WDM sample, with the blue and orange lines showing the least-squares fit for each sample, respectively. The observational datasets are the same as in \protect\ref{fig:dust-to-stellar-mass}.
    The black dashed line shows the fit to observational data from \protect\cite{dacunha_new_2010}.}
    \label{fig:dust-to-SFR}
\end{figure}

The dust mass shows a significant correlation with SFR, which matches moderately well with observed trends from \cite{dacunha_new_2010}, and with the same subset of observations as above, in \ref{fig:dust-to-SFR}. We apply a simple power-law fit of the form
\begin{equation}
    M_{\text{}dust} = \alpha\times10^7 * \psi^\beta,
\end{equation}
where $\psi$ is the integrated SFR for each halo, and $\alpha$ and $\beta$ are free parameters fitted to the data using a least-squares method. This resulted in similar fits for both DM models, with the CDM sample best-fit by $\alpha=3.13 \pm 0.57$ and $\beta = 0.65 \pm 0.10$, compared to the WDM sample where $\alpha=2.38 \pm 0.45$ and $\beta = 0.72\pm 0.09$. Both differ slightly from the \cite{dacunha_new_2010} fit but match well with the scatter in the subset of the other observational datasets. The FF-test results for the comparison of the SFR to dust mass have $p$-values again of 0.9 at $z=0$, providing no conclusive evidence that these distributions are drawn from different populations. 

As with the dust mass ratio, the difference between samples becomes slightly more significant at higher redshifts. For example, at $z=0.25$, the observable scaling relations are also noticeably different. At this redshift we see a greater separation in both the dust-to-stellar mass function and the dust-to-SFR function, though neither achieves the commonly accepted $p$-value of $<$0.05, to indicate that the two samples are likely drawn from different underlying populations\footnote{We note here that simply finding a $p$-value of $<$0.05 does not necessarily prove that the samples are unique, nor does a value $>$0.05 prove the samples are from the same distribution, just that the test was inconclusive.}. Given the relatively small sample size of 24 comparison simulations, these significance tests should not be treated as a statement of fact that the samples are different, but rather as an indication of where differences between DM models are more likely to arise.

\subsubsection{Alternative dust prescriptions}

So far, we have presented the results using the \dtm{} prescription, perhaps the most straightforward method for assigning dust masses to gas particles. While there are many more complex ways to model dust masses using simulated gas properties, here we present the dust-to-gas and best-fit dust prescriptions available to us in \powderday{}. Both these prescriptions produce a general increase in the total dust mass and a greater separation between the CDM and WDM samples. The median dust mass in the CDM sample using the dust-to-gas method is $1.2\times10^8$\msun{}, and $3.6\times10^8$\msun{} for the best-fit method, which represents increases of $2.1\times$ and $6.1\times$ compared to the \dtm{} method. The increase in median WDM dust mass is slightly lower, at only $1.7\times$ and $4.7\times$ for the dust-to-gas and best-fit methods, respectively, compared to the \dtm{} method. As such, the median ratio between the CDM and WDM samples at $z=0$ increases with changing dust production method, with a median $\sim1.3$ for both the dust-to-gas and best-fit methods. The evolution of the median ratio is similar across all dust production methods, although the peak difference occurs at slightly lower redshift, at $z\approx0.17$ for both the dust-to-gas and best-fit methods, rather than $z\approx0.25$ for the \dtm{} method.

The $z=0$ scaling relations are affected similarly by changing the dust prescription. With the increase in dust mass, we find a similar increase in the split between the two samples in both the dust-to-stellar mass and dust-to-SFR functions. It does not reach a point of statistical significance; the $p$-values are much lower for both the dust-to-gas and best-fit methods, but are still greater than 0.05 using the FF-test, and so we cannot meaningfully claim that the samples are drawn from unique distributions. 

However, at redshifts where the difference in dust mass peaks, we find that the scaling relations can produce statistically significant differences. The KS test of the dust distributions, used to assess whether the two are drawn from the same population, results in $p$-values of 0.01 and 0.004 for the DTG and best-fit methods, respectively, at $z=0.17$. This shows that there is evidence to reject the null hypothesis that dust distributions themselves come from the same underlying population. The FF-test results of the scaling relations are similarly impacted by the change in redshift, where the dust-to-stellar mass comparison yields $p$-values of 0.09 and 0.03, and the dust-to-SFR function comparison values of 0.07 and 0.03, again for the dust-to-gas and best-fit methods, respectively. 

It should be noted that neither DM model fits the observational relations better than the other, nor does either model match the observations shown here particularly well. Indeed, the FF-test of the CDM or WDM sample, compared to the combined observational sample for the dust-to-stellar mass function, indicates that the simulated samples are likely drawn from separate underlying distributions. The key driver of this is a very narrow range of allowed masses for resimulation targets in the zoom-in selection process compared to a much broader observational sample. We use this comparison to simply highlight where and how differences in dark matter models may arise. To this end, we may draw some tentative conclusions from this analysis. Firstly, using more sophisticated dust mass prescriptions yields different total masses but similar trends. Secondly, at higher redshifts, differences in dust properties between CDM and WDM haloes produce statistically significant observable differences in total dust mass and dust scaling relations, albeit with limited physical meaning given the previously stated caveats of simple modelling and small samples. Future work will focus on extending the sample to cover a broad range of haloes, in terms of both mass and environment. Additionally, we plan to update to more physically motivated baryonic models and WDM masses, though the use of more nuanced models will require more careful tuning, and an increase in WDM mass will lessen the amount of required tuning, combining to make disentangling differences caused by the dark matter model more complex.


\section{Discussion}\label{sec:Disc}
In this section, we discuss the previous results in greater detail, followed by a comparison to results from U24. 

\subsection{Dust masses}\label{sec:discussion_dust}

The dust masses across all simulations span roughly an order of magnitude, from 1.1$\times10^7$ to 1.8$\times10^8$ \msun{} using the \dtm{} production method. While this may seem like a large spread for such a narrow range of halo and stellar masses, it is generally in line with the spread from analyses of different datasets such as those discussed in \cite{dacunha_new_2010}, \cite{cortese_dust_2012}, and \cite{clemens_dust_2013}. We find our median dust masses, 5.8 (4.9) $\times10^7$\msun{} for the CDM (WDM) sample, is similar to the dust mass of M31 as a sense-check, at $5.4\times10^7$\msun{} \citep{draine_andromedas_2014}, although roughly an order of magnitude less than the dust mass in the sample of Milky Way-like simulations described in \cite{mckinnon_dust_2016}. The \cite{remy-ruyer_gastodust_2014} dust-to-gas production method and \cite{li_dusttogas_2019} best-fit parameters of their machine learning model both shift the median masses above $10^8$\msun{}, as shown in \ref{tab:suite_rr_dust_masses} and \ref{tab:suite_li_bf_dust_masses}, which is more in line with the \cite{mckinnon_dust_2016} simulations. Overall, the dust mass in a given simulation is highly stochastic, with substantial variability from one snapshot to the next, leading to large errors in the confidence intervals for median trends. Despite the apparent ensemble preference for dustier CDM galaxies, this highlights the difficulty of interpreting the results and reduces the efficacy of dust as a discriminant statistic.

The dust distributions are highly concentrated across all haloes, with almost all the dust residing in the inner 25 kpc. This is consistent with the expectation that the majority of the halo dust lies within the galaxy's disc, which is also seen in the simulations of \cite{mckinnon_dust_2016} and \cite{mckinnon_simulating_2017}. However, the concentrations found in this sample are perhaps higher than observations might suggest. Indeed, \cite{menard_measuring_2010} finds a halo dust mass of $\sim5\times10^7$\msun{} for $L^*$ galaxies, and \cite{peek_dust_2015} find a circumgalactic medium dust mass of $\sim6\times10^7$\msun{} for galaxies with stellar masses of $\sim6\times10^{10}$\msun{}. In this suite of simulations, the dust mass in the outer regions is often $\sim10^6$\msun{}, even for stellar masses of $\sim10^{11}$\msun{}. The lack of a stellar winds component in the subgrid model is a significant factor here, as the gas particles more readily sink and stay in the centre of the gravitational potential well, hence the high contrast in dust mass between central and outer regions. Winds and shocks from supernova also have a significant impact on the dust destruction rate \citep[e.g.][]{menon_bursty_2024, calura_interstellar_2025}. Including wind subgrid models would likely help smooth the dust distribution and reduce the stochasticity of the Gini scores, as the scores would be less dominated by satellites or the central galaxy than by the intrahalo medium.

As previously mentioned, the star formation model and dust production methods are simplified in efforts to disentangle the changes caused by altering the underlying dark matter and subgrid physics. Moving to a more advanced star formation and feedback model, and including self-consistent dust modelling, will have a large impact on dust properties. The dust trends shown here appear largely consistent within a single simulation pair, suggesting that a CDM universe may produce more dust than a WDM universe, but this requires further testing with more advanced models to confirm this trend. We do, however, expect that a full exploration of the dust mass evolution in this future work will prove suitable for comparison to observations. Properties such as total dust mass, stellar masses, and SFRs are readily available out to redshifts of $z>2$ \citep[e.g.][]{dunne_censusmetalshigh_2003, santini_evolution_2014, dacunha_alma_2015}, noting that studies at higher redshifts are biased towards more massive galaxies.


\subsubsection{Scaling relations}

We see a similar, moderate but limited, separation between the two dark matter samples when comparing them with the observational sample. The dust-to-stellar mass function, in \ref{fig:dust-to-stellar-mass}, indicates that, for the \dtm{} method, our samples both tend to under-produce dust in relation to fits from \cite{skibba_emission_2011} and \cite{peeples_budget_2014}. The primary driver of this is most likely an overly efficient star formation model, with greater stellar masses expected for a given halo mass. The median stellar mass for both CDM and WDM samples is approximately $1\times10^{11}$\msun{}, moderately higher than the observed Milky Way stellar mass estimates of 5 - 6$\times10^{10}$\msun{} from \cite{bland-hawthorn_galaxy_2016}. Barring a few outliers, these haloes generally land above both \cite{behroozi_comprehensive_2010} and \cite{moster_galactic_2013} stellar mass - halo mass relations. The other dust production methods can compensate for the enhanced stellar mass by boosting dust mass with the inclusion of further gas particle properties. The dust-to-gas method tends to place the sample most in line with observational dust-to-stellar mass functions, while the best-fit model increases the dust mass again, generally moving the sample above these fits to observational data.

The dust mass shows a much clearer correlation with the SFR of each galaxy, despite the narrow range of halo masses, in \ref{fig:dust-to-SFR}. Here, the \dtm{} method most closely matches the observations described in \cite{clemens_dust_2013}, with a consistent spread in dust masses. The fits to the dust mass using the dust-to-gas method also match observations reasonably well, although they are slightly higher; the best-fit method is furthest from the observational fits to the dust-to-SFR function. 

These two scaling relations were chosen because of their common use in observational samples and ease of analysis in simulations. At certain epochs, we find a statistically significant separation between the two simulated samples in both more complex dust production methods. However, this is not universally seen in all production methods, and any signal is mostly lost by $z=0$. While there is some indication that dust can be used in dark matter research, the sample presented here is not sufficiently large to strongly claim evidence of clear separation between the samples, and is biased to only galaxies of Milky Way mass. Future work should focus on generating a larger statistical sample, ideally hundreds of galaxies across a broad mass spectrum, which would help characterise trends with mass and evolution. 





\subsection{Limitations of the Gini score metric}\label{sec:discussion_gini}

Contrary to previous results, the analysis of the suite of simulations generally shows little difference in the mean Gini scores of the cold and warm dark matter simulations. With the larger sample of simulations, it is evident that the presence of satellites dominates the Gini score. Because the central galaxy is masked, the relative difference in dust surface density between an infalling satellite and the diffuse halo component is significant, and large satellites will increase and dominate the Gini score. As such, the difference in Gini score driven by the change in the dark matter model will be difficult to uncover, as the large satellites are not as suppressed under a 3.5 keV WDM particle, ultimately hampering its usefulness as a discriminatory metric under the conditions examined in this work. 

Despite the mean difference being close to zero, changing the dark matter model causes large stochastic differences in a single simulation comparison, as seen in \ref{fig:gini_difference_all}. Infalling satellites, especially large satellites, have a significant effect on the Gini score. Consequently, if the same satellite merges into the halo at even very slightly different times, this can cause significant differences in the scores between the simulations. Additionally, the different concentrations found in cold and warm dark matter haloes will affect how long a satellite spends in the halo before being disrupted and merging into the central region \citep{mao_dependence_2015}. As cold dark matter haloes typically have higher concentrations, the merging timescale also has a significant impact on the evolution of the Gini scores between a simulation pair. Differences in both infall times and merger times occur in such a manner that the difference between the simulations can swing from large positive to negative values, resulting in a near-zero mean difference with a large variance. 

When studying the difference between the two cold dark matter simulations, \ref{fig:gini_difference_dtm_cdm_sn_005} shows that a reduction in the supernova efficiency parameter does induce a slight reduction in Gini scores, predominantly at higher redshift. This is likely caused by marginally lower dust masses in satellite galaxies, with similar diffuse components. The difference in Gini scores between the two simulations with 10\% efficiency, in \ref{fig:gini_difference_dtm_wdm_sn_010}, shows a very slight increase in Gini scores in warm dark matter over cold dark matter simulations. This matches well with the fact that the dust masses tend to be slightly higher in the outer regions of the warm dark matter simulations. These two plots show that the Gini scores exhibit a slight sensitivity to the baryonic model being used, but the differences in halo formation and structure caused by the different dark matter models dominate the variance. 

This sample illustrates the limitations of using the Gini score in cases like this. There are hints that the different baryonic models and dark matter models do generate differences, but the scatter is so large as to make drawing conclusions difficult. To uncover any signal in the Gini score would require several extensions, namely, heavily broadening the sample and working to recover structural information in the centre of the halo. The combination of changes in the dark matter model and the baryonic model does generate differences in the Gini scores, but any underlying behaviour is not well captured in only 24 comparison simulations, which would perhaps be more evident or well characterised in a sample of $\sim$100's of galaxies. Updating the baryonic model would help recreate more accurate galactic discs, allowing us to characterise any differences in both the inner and outer regions of the haloes. Although, as previously mentioned, more complex baryonic models would require more care in the tuning and analysis of differences that result from the baryonic model as opposed to changes in the dark matter model.

\subsection{Comparison with findings from U24}\label{sec:discussion_pap_1}

In the analysis of U24, it was initially indicated that the different dark matter models would produce significant differences in both the dust properties and the Gini scores, a prediction we do not find in this sample. Here, we discuss why those simulations differed significantly from the rest of the simulations in the suite. 

\begin{figure*}
    \begin{subfigure}[]{0.48\textwidth}
        \centering
        \includegraphics[width=\textwidth]{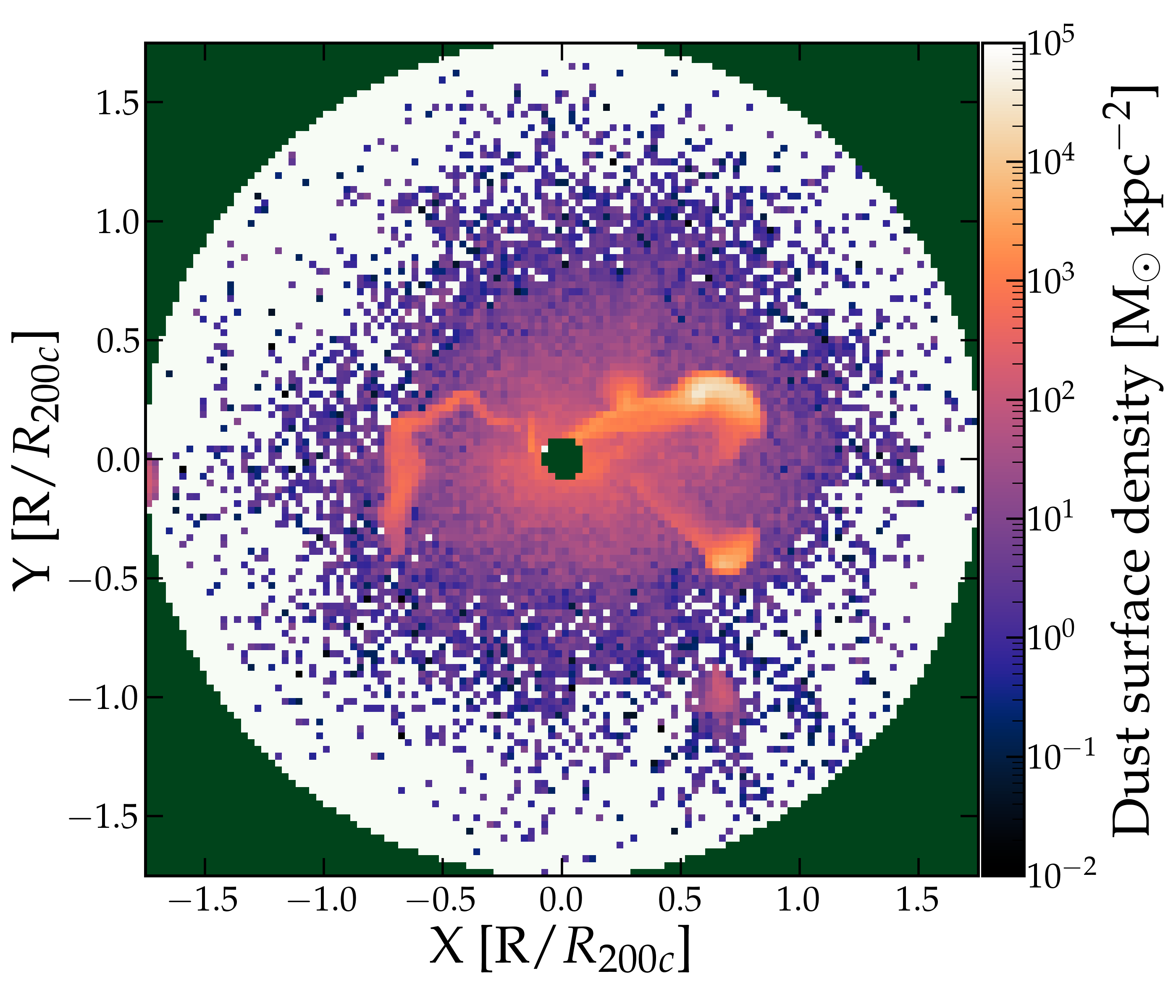}
        \caption{CDM}
        \label{fig:pap_3_disc_cdm_dust_dist}
    \end{subfigure}
    \hfill
    \begin{subfigure}[]{0.48\textwidth}
    \centering
        \includegraphics[width=\textwidth]{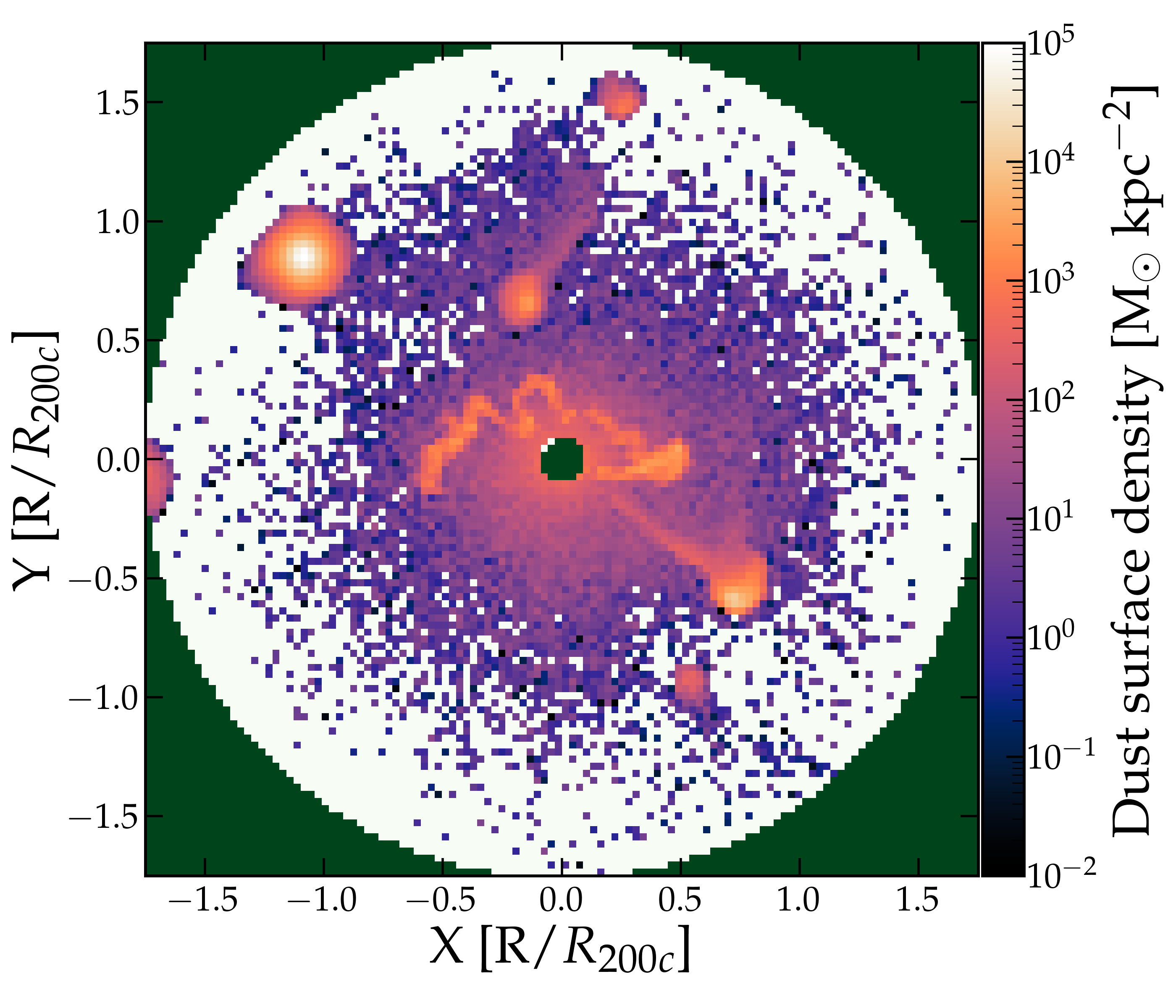}
        \caption{WDM}
        \label{fig:pap_3_disc_wdm_dust_dist}
    \end{subfigure}
    \hfill
    \caption{A reanalysis of the dust surface density of the simulation used in U24, including bins out to a radius of 1.75$\times$\rthc{}. White bins are empty but still included in the Gini score calculations; the green regions, a circle with a 15kpc radius over the central galaxy and beyond 1.75$\times$\rthc{} are excluded from the calculation.}  
    \label{fig:pap_1_redo}
\end{figure*}

The simulations used in U24 had very similar satellite mass functions, especially at the high-mass end, and both matched well with the observed Milky Way satellites. However, the positioning of the largest satellite differed significantly in each simulation. Due to a delay in the infall time, the largest satellite in the warm dark matter simulation, while identified and considered gravitationally bound by \textsc{subfind-hbt}, was still on first infall and located just beyond \rthc{}. This can be seen in \ref{fig:pap_1_redo}, where the radius shown has been increased to 1.75$\times$\rthc{}. The same satellite\footnote{As determined based on the shared dark matter particles in both simulations} in the cold dark matter halo is located to the right of the central galaxy,  well within \rthc{}. The difference in the satellite position is a key driver of the different results found in U24 compared to the ensemble trends in dust mass and Gini score found in this work.

In terms of the dust mass ratio, the U24 simulations stand out compared to the suite average, with the largest mass ratio at 4.35. When calculating the dust mass with \powderday{}, we assign masses for every gas particle within 500kpc from the halo centre. However, we only use particles within \rthc{} for all dust metrics, but overlook bound structures beyond this radius. When the total dust mass is calculated for every gas particle within the 500kpc limit, the ratio between the two dark matter models is reduced but still heavily favours the cold dark matter simulation. The total mass in the cold dark matter simulation increases very slightly, from $4.6\times10^7$ to $4.7\times10^7$\msun{}, while the warm dark matter has a slightly larger increase, from $1.1\times10^7$ to $1.7\times10^7$\msun{}, bringing the dust mass ratio down from 4.3 to 2.7. The drop in dust mass ratio is caused by now including the large satellite; however, a dust mass ratio of 2.7 would still be the second largest among any simulation in \ref{tab:suite_dtm_dust_masses}. As such, excluding the dust mass from the satellite doesn't have a significant impact on the overall trend, only on its amplitude.

\begin{figure}
    \begin{subfigure}[]{\columnwidth}
        \centering
        \includegraphics[width=\textwidth]{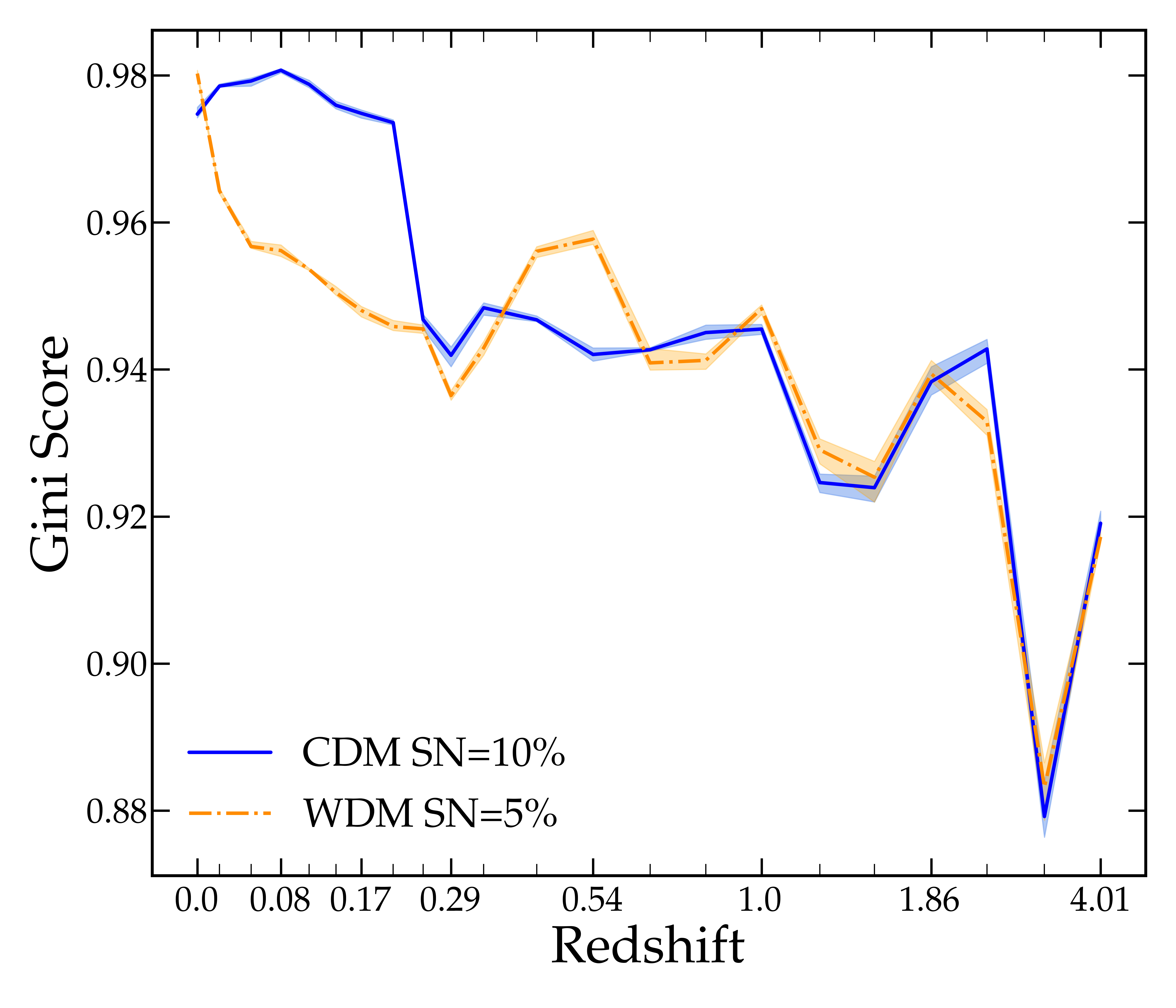}
        \caption{Scores}
        \label{fig:pap_3_disc_gini_score}
    \end{subfigure}
    \hfill
    \begin{subfigure}[]{\columnwidth}
    \centering
        \includegraphics[width=\textwidth]{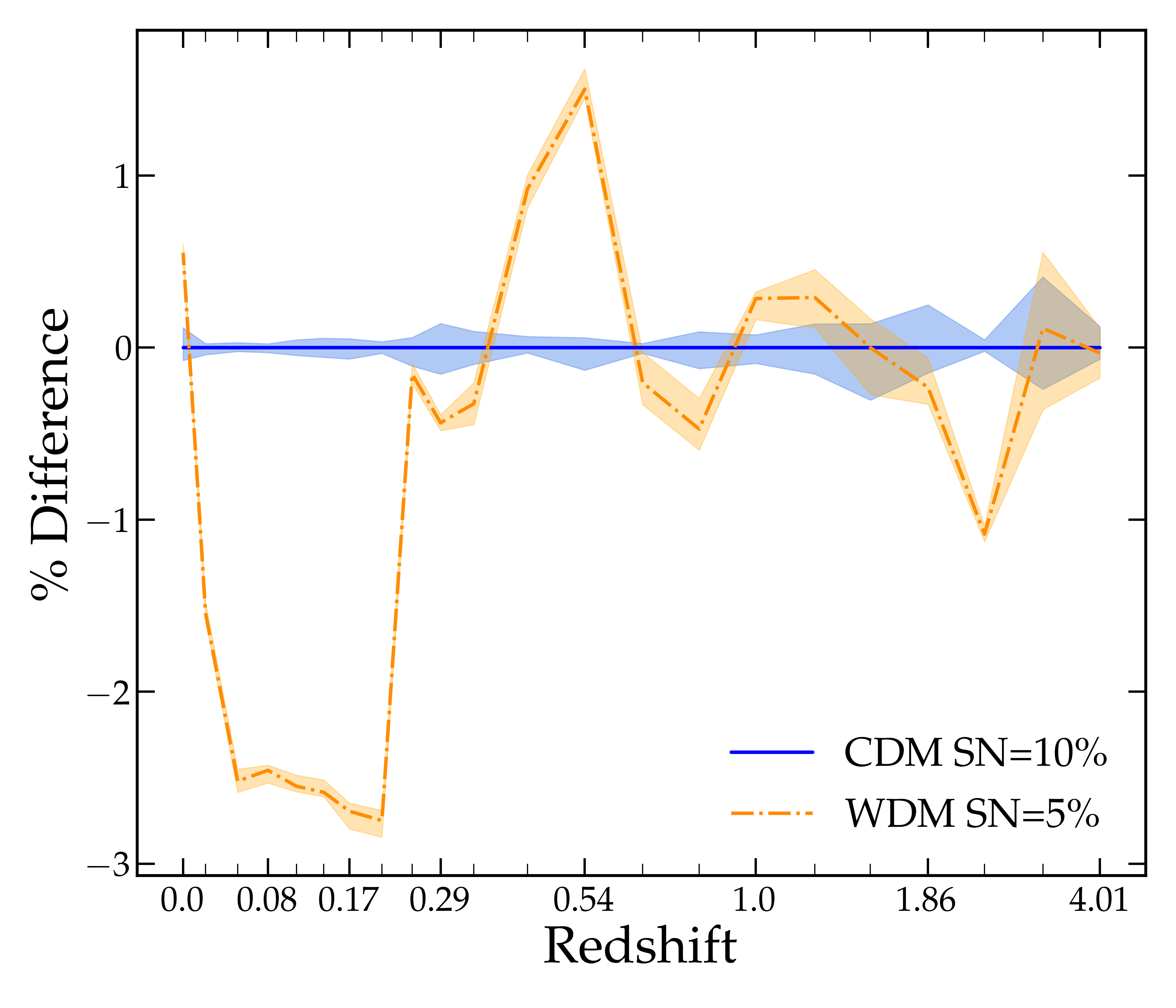}
        \caption{Difference}
        \label{fig:pap_3_disc_gini_diff}
    \end{subfigure}
    \hfill
    \caption{A reanalysis of the Gini scores from the simulation used in U24, with the scores shown in the top plot and the relative difference in the lower plot. The large difference at low redshift is caused by the different merger times of a large satellite galaxy.}
    \label{fig:pap_1_redo_gini}
\end{figure}

The Gini scores are more susceptible to change when the radius is increased to include this satellite. We repeat the Gini analysis from U24, but increase the radius to 1.75$\times$\rthc{}, and we show the results in \ref{fig:pap_1_redo_gini}. With the increased radius we see in \ref{fig:pap_3_disc_gini_score} that the Gini score for the warm dark matter simulation increases back up to the same level as the cold dark matter simulation, with the differences also roughly at parity in \ref{fig:pap_3_disc_gini_diff} by $z=0$. This shows that the total dust is similarly concentrated in the large satellites between both dark matter models; it was just the difference in satellite position driving the separation. Though the scores are similar after the radius increase, the difference between $z\simeq0.2$ and $z=0$ is a good indicator of the structural differences between the two dark matter models.

Although it is unfortunate that the window affects the Gini score to such an extent, some window must be chosen. We chose \rthc{}, given its wide use in defining an edge for simulated haloes. For future studies that utilise the Gini coefficient, adopting an observer-based approach could be beneficial. For example, using the stellar mass-size relation for late-type galaxies \citep[Equation 18 from][]{shen_size_2003} and multiplying the derived half light radius $\overline{R}_{50}$ by an arbitrary factor of 50 recovers a reasonable approximation of \rthc{}. Ultimately, this doesn't fix the potential change in dust properties caused by differences in satellite position between dark matter models, but it at least provides an observational basis for the window of study rather than a purely simulation-based metric.

\section{Conclusions}\label{sec:Conc}
In this paper, we presented a suite of 24 cold and warm dark matter simulation pairs to extend the work described in \cite{ussing_using_2024}. Based on that work, the expectation was that a cold dark matter hydrodynamical simulation with an increased supernova rate would generate more dust in a particular galaxy than a warm dark matter simulation of the same galaxy with a reduced supernova efficiency. Additionally, this dust would also be more structurally concentrated in the cold dark matter simulations, calculated using the Gini coefficient with \ref{eq:gini}.

Each of the 24 simulation pairs was created using the same random seed, using haloes selected from a dark matter only parent simulation. Resimulation targets were identified as Milky Way sized, with a mass \mthc{} between $0.55-2.62\times10^{12}$\msun{} \citep{bland-hawthorn_galaxy_2016}, and no large neighbouring haloes. The particles were then traced back to the initial redshift to define the Lagrangian region, which was then resimulated at higher resolution with hydrodynamic elements. The simulations were then run to $z=0$ using two different baryonic prescriptions: the fiducial supernova efficiency of 10 per cent as defined in \cite{springel_cosmological_2003}, and a reduced efficiency of 5 per cent. We nominated the CDM simulation suite with the 10 per cent efficiency as our fiducial CDM sample and the WDM suite with the reduced efficiency of 5 per cent as our fiducial WDM sample.

The simulations were subsequently post-processed with \powderday{} \citep{narayanan_powderday_2021} to predict the dust properties of the resimulated Milky Way analogues, using multiple dust prescriptions: a \dtm{} ratio of 0.4 \citep{dwek_evolution_1998}, the dust-to-gas relation from \cite{remy-ruyer_gastodust_2014}, and the \cite{li_dusttogas_2019} best-fit parametrisation of their machine learning model. We analysed the data in the context of the ensemble trends, followed by a comparison with previous work.

The key results of this work are summarised as follows:

\begin{itemize}
    \item The baryonic tuning between the cold and warm dark matter simulations in the \gadget{} suite produced a range of satellite functions with similar means. Satellites with stellar masses in the range of the Large and Small Magellanic Clouds were uncommon in this sample, as shown in \ref{fig:satellite_pop_all}. 
    
    \item When analysing the total dust masses within \rthc{} for our fiducial samples, we saw a mean increase of 30 per cent and a median increase of 6 per cent CDM dust mass compared to WDM at $z=0$, shown in \ref{tab:suite_dtm_dust_masses}. The dust masses were also highly concentrated in the central regions of the haloes, with an average of $\sim$90 per cent of the dust located in the inner 25 kpc of the halo in the CDM sample compared to 85 per cent in the WDM sample.
    
    \item The total dust mass shows a preference for higher dust masses in the CDM sample across significant redshift evolution, in \ref{fig:dust_mass_ratio_evo}. The median difference ranged between $\sim$5-25 per cent across redshifts $z\approx2$ to $z=0$, peaking around $z=0.25$ for the \dtm{} method, albeit with significant errors given by bootstrapped confidence intervals. The dust-to-gas and best-fit methods had a greater increase in the median ratio, both with a 30 per cent increase at $z=0$ and peaking slightly later at $z\approx.17$ with a median increase of CDM to WDM dust mass of 50 per cent.

    \item The ensemble Gini scores showed large stochastic differences between the two dark matter models for each simulation pair, with little identifiable trend in the mean difference, shown in \ref{fig:gini_difference_all}. There was a slight reduction in the dust concentrations at high redshift in the warm dark matter simulations, but the 99\% confidence interval was still inclusive of zero difference.
    
    \item The individual scores in the warm and cold dark matter simulations both shared a rapid consolidation phase between $z\simeq4$ and $z\simeq1$, followed by a divergence of scores and a slight downward trend in the mean, shown in \ref{fig:gini_score_all}. The scores at lower redshifts were dominated by infalling haloes and exhibited a large spread in scores.

    \item Simple statistical tests of the dust masses and common scaling relations, the dust-to-stellar mass and dust-to-SFR functions, showed that at $z=0$, the CDM and WDM samples do not produce statistically significant differences for any dust production metric. 

    \item At higher redshifts where differences peak, the \cite{li_dusttogas_2019} best-fit dust production model produces distributions that do constitute statistical significance in a KS-test and FF-tests of the dust-to-stellar mass and dust-to-SFR distributions. However, this is not found across all dust production methods and only in a very narrow redshift range.
    

\end{itemize}


This work has shown that, while some of the initial findings of U24 are apparent in a larger suite, separating dark matter models by dust properties requires further study beyond the suite presented here. The total dust mass, and dust correlations, can produce separations that may be observable, but the limitations of the dust modelling and sample size hamper the use of dust in determining an observational preference for one dark matter model over another. The structural properties also proved highly stochastic, rendering the Gini metric ineffective at distinguishing between dark matter models. 






Despite the lack of clear delineation between dark matter models, the hints of significant differences at higher redshifts warrant further investigation. Future work will focus on improved baryonic modelling and a large sample of galaxies covering a wider mass range. The advanced modelling will provide a more realistic comparison point for observed scaling relations, with enough galaxies to make quantitative conclusions in preference for one dark matter model over another. 


\section*{Acknowledgements}
The authors would like to thank the referee for their valuable comments in improving this publication. This research was supported in part by the Australian Government through the Australian Research Council Centre of Excellence for Dark Matter Particle Physics (CDM, CE200100008), and in part by the Australian Research Council Centre of Excellence for All Sky Astrophysics in 3 Dimensions (ASTRO 3D, CE170100013). AU acknowledges the support of an Australian Government Research Training Program Scholarship, and DC acknowledges the support of an ARC Future Fellowship (FT220100841). This work was performed on the OzSTAR national facility at Swinburne University of Technology. The OzSTAR program receives funding in part from the Astronomy National Collaborative Research Infrastructure Strategy (NCRIS) allocation provided by the Australian Government, and from the Victorian Higher Education State Investment Fund (VHESIF) provided by the Victorian Government.

\section*{Data Availability}
The simulations used in this study are available and accessible on reasonable request. The data products of this work can be shared upon request to the corresponding author.



\bibliographystyle{mnras}
\bibliography{references} 

\appendix

\section{Alternative dust production mechanisms}\label{app:appendix_A}
Here, we present a summary of the results of the dust analysis using the alternate dust production mechanisms from \powderday{}. In \ref{tab:suite_rr_dust_masses} and \ref{tab:suite_li_bf_dust_masses}, we show the total dust mass using the \cite{remy-ruyer_gastodust_2014} and \cite{li_dusttogas_2019} best-fit model, respectively. As in U24, these two mechanisms show significant increases in dust mass over the \dtm{} ratio. The DTG ratio serves as a kind of midpoint between the other two dust production mechanisms in terms of dust mass produced, showing only a $\sim2\times$ increase in total dust over the \dtm{} ratio, compared to the $\sim6\times$ increase in the best-fit parameters. The median and mean dust mass ratio also increase, both to $\sim1.3$, importantly retaining the preference for greater dust in the CDM sample.

\begin{table}
    \centering
    \begin{tabular}{c|c|c|c}
        \hline
        Simulation  & CDM 10\%  & WDM 5\% & CDM/WDM\\
        ID          & ($\times10^7$\msun{}) & ($\times10^7$\msun{}) &  mass ratio\\
        \hline
        ID-00 & 36.07 & 28.72  & 1.256  \\
        ID-01 & 26.54 & 16.6  & 1.599  \\
        ID-02 & 19.43 & 14.5  & 1.34  \\
        ID-03 & 32.09 & 31.47  & 1.019  \\
        ID-04 & 2.84 & 2.93  & 0.967  \\
        ID-05 & 26.58 & 1.85  & 14.356  \\
        ID-06 & 27.75 & 38.58  & 0.719  \\
        ID-07 & 10.88 & 7.73  & 1.407  \\
        ID-08 & 10.65 & 6.0  & 1.777  \\
        ID-09 & 13.81 & 10.41  & 1.327  \\
        ID-10 & 0.66 & 0.94  & 0.708  \\
        ID-11 & 10.45 & 10.2  & 1.025  \\
        ID-12 & 0.21 & 7.07  & 0.03  \\
        ID-13 & 8.05 & 6.76  & 1.191  \\
        ID-14 & 7.53 & 6.8  & 1.107  \\
        ID-17 & 13.31 & 8.24  & 1.615  \\
        ID-28 & 4.22 & 1.19  & 3.552  \\
        ID-34 & 5.55 & 8.25  & 0.672  \\
        ID-46 & 11.17 & 1.33  & 8.369  \\
        ID-57 & 5.22 & 3.01  & 1.738  \\
        ID-61 & 21.25 & 12.45  & 1.706  \\
        ID-70 & 16.82 & 11.77  & 1.429  \\
        ID-80 & 16.57 & 12.12  & 1.367  \\
        ID-93 & 16.63 & 13.92  & 1.194  \\
        \hline
        Median & 12.241 & 8.248 & 1.333 \\
        Mean & 14.345 & 10.952 & 2.145 \\
        \hline
    \end{tabular}
    \caption{The dust masses for each simulation within \protect\rthc{}, using the \protect\cite{remy-ruyer_gastodust_2014} dust-to-gas model.}
    \label{tab:suite_rr_dust_masses}
\end{table}

\begin{table}
    \centering
    \begin{tabular}{c|c|c|c}
        \hline
        Simulation  & CDM 10\%  & WDM 5\% & CDM/WDM\\
        ID          & ($\times10^7$\msun{}) & ($\times10^7$\msun{}) &  mass ratio\\
        \hline
        ID-00 & 98.43 & 85.62  & 1.15  \\
        ID-01 & 73.86 & 43.76  & 1.688  \\
        ID-02 & 45.01 & 33.11  & 1.359  \\
        ID-03 & 69.12 & 76.89  & 0.899  \\
        ID-04 & 7.23 & 7.62  & 0.948  \\
        ID-05 & 75.63 & 4.56  & 16.594  \\
        ID-06 & 74.28 & 91.09  & 0.815  \\
        ID-07 & 34.37 & 23.89  & 1.439  \\
        ID-08 & 30.51 & 15.32  & 1.991  \\
        ID-09 & 37.27 & 28.95  & 1.288  \\
        ID-10 & 0.49 & 0.75  & 0.653  \\
        ID-11 & 27.38 & 26.38  & 1.038  \\
        ID-12 & 0.13 & 18.55  & 0.007  \\
        ID-13 & 21.55 & 17.59  & 1.225  \\
        ID-14 & 19.79 & 17.48  & 1.132  \\
        ID-17 & 36.63 & 20.1  & 1.823  \\
        ID-28 & 13.18 & 2.84  & 4.635  \\
        ID-34 & 18.47 & 20.53  & 0.9  \\
        ID-46 & 29.2 & 3.32  & 8.797  \\
        ID-57 & 33.39 & 22.02  & 1.517  \\
        ID-61 & 60.53 & 36.56  & 1.656  \\
        ID-70 & 42.22 & 28.8  & 1.466  \\
        ID-80 & 42.7 & 29.3  & 1.458  \\
        ID-93 & 42.65 & 34.88  & 1.223  \\
        \hline
        Median & 35.501 & 22.952 & 1.323\\
        Mean & 38.917 & 28.746 & 2.3207 \\
        \hline
    \end{tabular}
    \caption{The dust masses for each simulation within \protect\rthc{}, using the \protect\cite{li_dusttogas_2019} best-fit model.}
    \label{tab:suite_li_bf_dust_masses}
\end{table}

\section{Non-fiducial Gini analysis }\label{app:appendix_B}
\begin{figure*}
    \hfill
    \begin{subfigure}{\columnwidth}
    \centering
        \includegraphics[width=\textwidth]{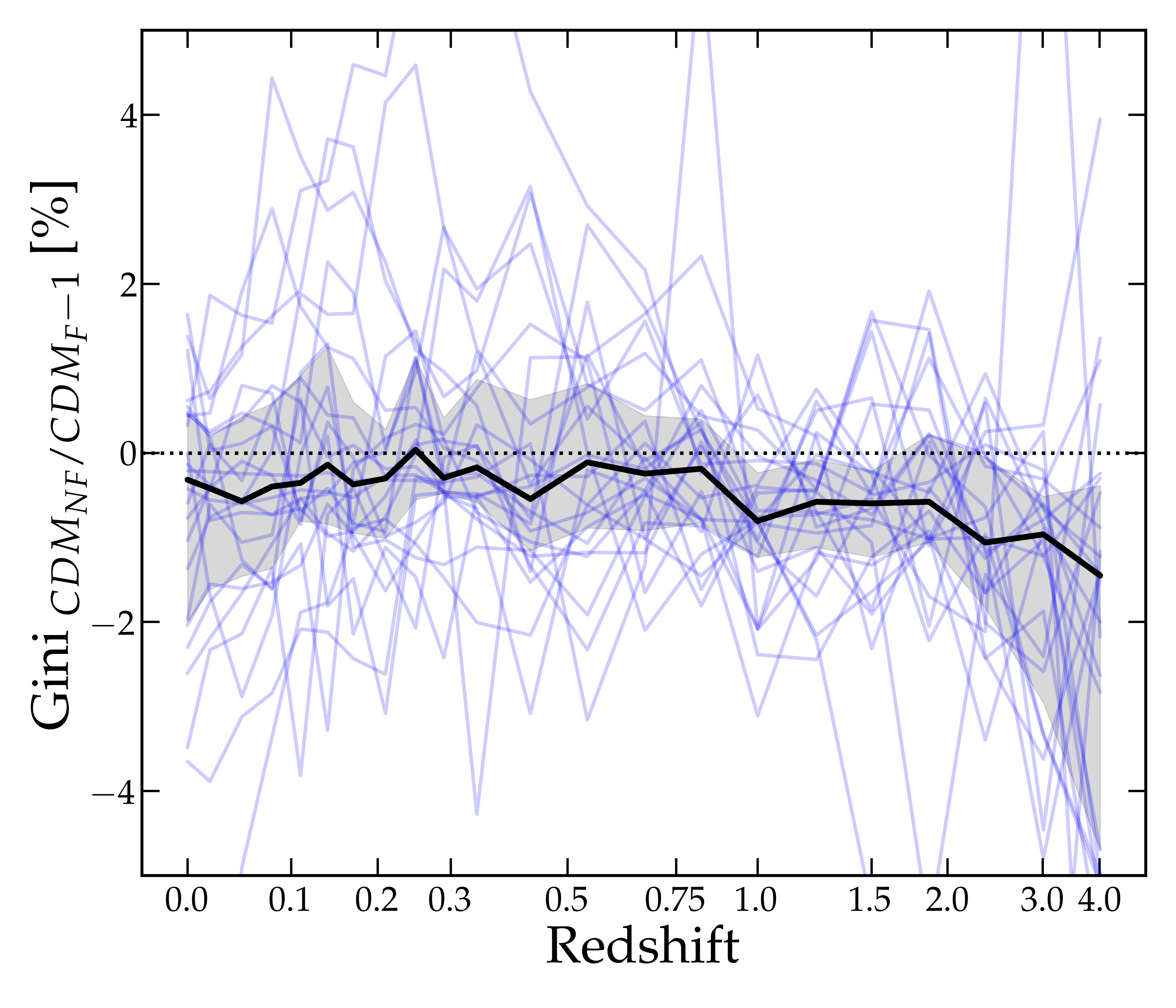}
        \caption{CDM 5\% to CDM 10\% simulation comparison}
        \label{fig:gini_difference_dtm_cdm_sn_005}
    \end{subfigure}
    \hfill
    \begin{subfigure}{\columnwidth}
    \centering
        \includegraphics[width=\textwidth]{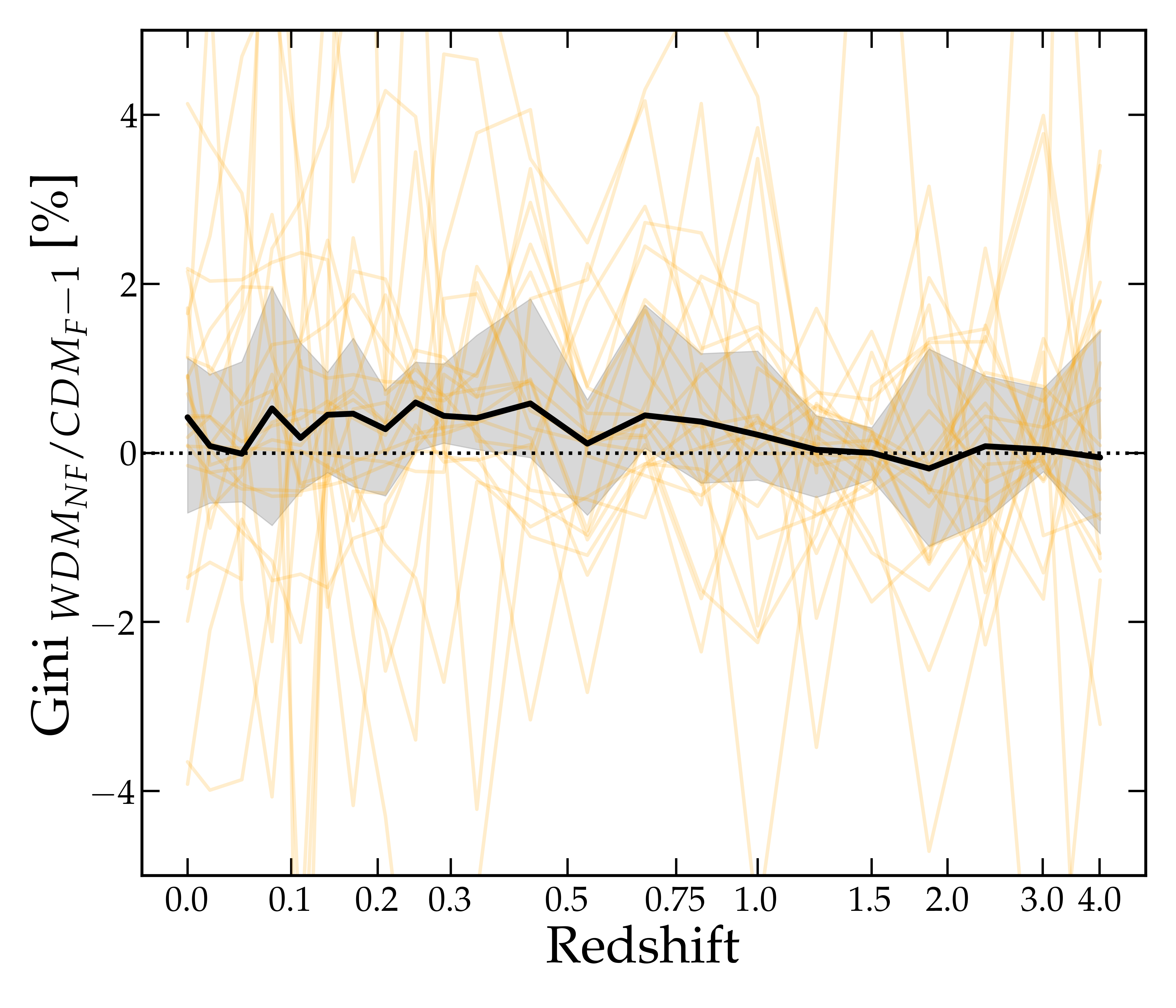}
        \caption{WDM 10\% to CDM 10\% simulation comparison}
        \label{fig:gini_difference_dtm_wdm_sn_010}
    \end{subfigure}
    \caption{The difference between the Gini scores in the two non-fiducial cold and warm dark matter simulations compared with the fiducial cold dark matter simulation, as denoted by the NF and F subscripts, respectively. The left-hand plot shows the difference between the two sets of CDM simulations, or the difference caused solely by changing the supernova efficiency. The right-hand plot shows the difference between the cold and warm dark matter pairs, both using the fiducial 10 per cent supernova efficiency. The median and 99\% confidence interval are shown as the black line and grey shaded region, respectively, in both plots. Despite a large spread in differences, the median of the sample remains close to zero.}
    \label{fig:appendix_gini_difference_all}
\end{figure*}

Here we show the differences in the Gini scores between non-fiducial CDM and WDM simulations, compared to the fiducial CDM simulations, with dust masses calculated using the \dtm{} ratio. The comparison between the two cold dark matter simulations, shown in \ref{fig:gini_difference_dtm_cdm_sn_005}, shows a slight reduction in the mean difference, though still quite close to zero, with a much lower scatter than either of the comparisons to the warm dark matter simulations. In \ref{fig:gini_difference_dtm_wdm_sn_010}, the difference between the two simulations with 10\% supernova efficiencies, we see a slight increase in the Gini score in warm dark matter. This is likely due to the slight increase in dust mass outside the masking region as seen in \ref{tab:suite_dtm_outer_dust_masses}. These show that the Gini is sensitive to the baryonic parameters, but the scatter often dominates the signal, especially in comparison to the WDM simulations.

\bsp	
\label{lastpage}
\end{document}